\documentclass[sigconf]{acmart}
\usepackage{subfigure}
\usepackage{amsfonts}
\usepackage{bbding}
\usepackage{colortbl}
\usepackage{xcolor}
\usepackage{soul} 

\usepackage{multirow}
\usepackage{color}
\usepackage{amsmath}
\usepackage{hyperref}
\usepackage{balance}
\AtBeginDocument{%
  \providecommand\BibTeX{{%
    \normalfont B\kern-0.5em{\scshape i\kern-0.25em b}\kern-0.8em\TeX}}}

\copyrightyear{2024}
\acmYear{2024}
\setcopyright{acmlicensed}\acmConference[MM '24]{Proceedings of the 32nd
ACM International Conference on Multimedia}{October 28-November 1,
2024}{Melbourne, VIC, Australia}
\acmBooktitle{Proceedings of the 32nd ACM International Conference on
Multimedia (MM '24), October 28-November 1, 2024, Melbourne, VIC, Australia}
\acmDOI{10.1145/3664647.3681643}
\acmISBN{979-8-4007-0686-8/24/10}

\begin{document}

\title{HINER: Neural Representation for Hyperspectral Image}


\author{Junqi Shi}
\affiliation{%
  \institution{NanJing University}
  % \streetaddress{1 Th{\o}rv{\"a}ld Circle}
  \city{NanJing}
  \country{China}
  }
\email{junqishi@smail.nju.edu.cn}

\author{Mingyi Jiang}
\affiliation{%
  \institution{NanJing University}
  % \streetaddress{1 Th{\o}rv{\"a}ld Circle}
  \city{NanJing}
  \country{China}
  }
\email{jiangmy@smail.nju.edu.cn}

\author{Ming Lu}
\authornote{Corresponding author.}
\affiliation{%
  \institution{NanJing University}
  % \streetaddress{1 Th{\o}rv{\"a}ld Circle}
  \city{NanJing}
  \country{China}
  }
\email{minglu@nju.edu.cn}

\author{Tong Chen}
\affiliation{%
  \institution{NanJing University}
  % \streetaddress{1 Th{\o}rv{\"a}ld Circle}
  \city{NanJing}
  \country{China}
  }
\email{chentong@nju.edu.cn}

\author{Xun Cao}
\affiliation{%
  \institution{NanJing University}
  % \streetaddress{1 Th{\o}rv{\"a}ld Circle}
  \city{NanJing}
  \country{China}
  }
\email{caoxun@nju.edu.cn}

\author{Zhan Ma}
\affiliation{%
  \institution{NanJing University}
  % \streetaddress{1 Th{\o}rv{\"a}ld Circle}
  \city{NanJing}
  \country{China}
  }
\email{mazhan@nju.edu.cn}

% \affiliation{%
%   \institution{NanJing University}
%   % \streetaddress{1 Th{\o}rv{\"a}ld Circle}
%   \city{NanJing}
%   \country{China}
%   }

\renewcommand{\shortauthors}{Junqi Shi et al.}

\begin{abstract}
  % As a prevalent scientific data format with extensive applications, the efficient compression of hyperspectral images (HSI) and ensuring high-quality downstream tasks have garnered significant attention. This paper introduces {HINER}, a novel approach for compressing HSI using Neural Representation. 

  This paper introduces {HINER}, a novel neural representation for compressing HSI and ensuring high-quality downstream tasks on compressed HSI. 
  HINER fully exploits inter-spectral correlations by explicitly encoding of spectral wavelengths and achieves a compact representation of the input HSI sample through joint optimization with a learnable decoder. By additionally incorporating the Content Angle Mapper with the L1 loss, we can supervise the global and local information within each spectral band, thereby enhancing the overall reconstruction quality. For downstream classification on compressed HSI, we theoretically demonstrate the task accuracy is not only related to the classification loss but also to the reconstruction fidelity through a first-order expansion of the accuracy degradation, and accordingly adapt the reconstruction by introducing Adaptive Spectral Weighting. Owing to the monotonic mapping of HINER between wavelengths and spectral bands, we propose Implicit Spectral Interpolation for data augmentation by adding random variables to input wavelengths during classification model training. 
  Experimental results on various HSI datasets demonstrate the superior compression performance of our HINER compared to the existing learned methods and also the traditional codecs. Our model is lightweight and computationally efficient, which maintains high accuracy for downstream classification task even on decoded HSIs at high compression ratios. Our materials will be released at \url{https://github.com/Eric-qi/HINER}.
\end{abstract}






\maketitle

\section{Introduction}

The hyperspectral image (HSI)
%compared to RGB images that record three spectral (color) channels per pixel, 
uses tens of spectral bands across {a wide range} of electromagnetic wavelengths at each pixel position to capture the physical scene~\cite{varshney2004advanced}, by which it promises exceptional capabilities for tasks like object detection, material inspection, and scene analysis for applications in agriculture \cite{willoughby1996application}, aerospace industry \cite{grahn2007techniques}, remote sensing \cite{goetz1985imaging}, etc.  However, compared with the three-channel RGB image, orders of magnitude more spectral channels in each HSI sample present practical challenges for storage and transmission, largely impeding its use in various applications. As a result, efficient lossy HSI compression is highly desired. 

{In addition to traditional rules-based HSI compression methods using transform~\cite{karami2012compression} or linear prediction~\cite{mielikainen2003linear}, over the past few years, there has been a growing interest in leveraging deep learning techniques for HSI compression~\cite{guo2023hyperspectral, dua2020comprehensive, dua2021convolution, la2022hyperspectral, guo2021learned}. Through the powerful modeling capabilities of neural networks, learned HSI compression has demonstrated noticeable compression efficiency improvement. 
As one of them, implicit neural representations (INRs) have gained increasing popularity for representing natural signals with intricate characteristics.}
{The fundamental concept behind INR is to represent a signal as a tailored neural network, thus, the compression of the input signal is translated into the compression of the neural model itself. Such INR methods significantly diminish the requisite for extensive training data given the high cost of acquiring large-volume HSIs~\cite{zhang2024compressing}, and also streamline the decoding process. Zhang et al.~\cite{zhang2024compressing} and Rezasoltani et al.~\cite{rezasoltani2023hyperspectral, rezasoltani2023hyperspectral_v0} have pioneered the exploration of neural representation for HSI compression. Both of them directly adopt the architecture of SIREN~\cite{sitzmann2020implicit}, which employs a cascade of Multi-Layer Perceptions (MLP) with periodic activation functions, for pixel-by-pixel compression of the input HSIs. However, such a pixel-wise approach is built upon the assumption of spatial redundancy and represents HSIs through spatial position embedding, which disregards the strong correlation across spectral bands, leading to performance limitations. Even worse, signal distortion induced by those lossy compression methods notably deteriorates the accuracy of the downstream task (e.g., classification), which makes them extremely difficult to promote in applications.}

\begin{figure}[t]
\centering
    \subfigure[9-th frame]{
    \includegraphics[width=0.13\textwidth]{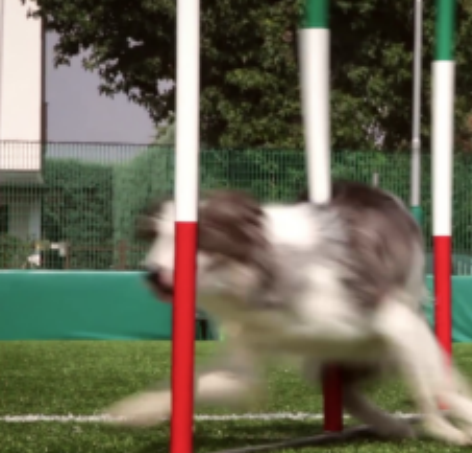}}
    \subfigure[10-th frame]{
    \includegraphics[width=0.13\textwidth]{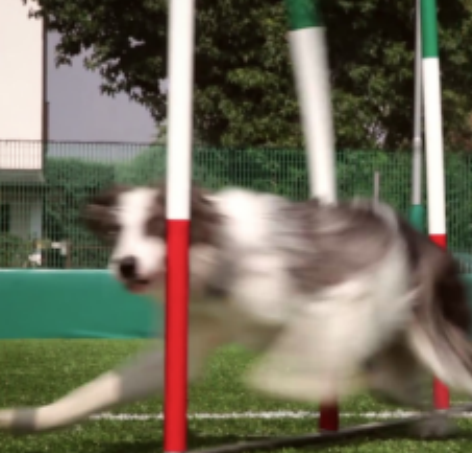}}
    \subfigure[diff. (motion)]{
    \includegraphics[width=0.13\textwidth]{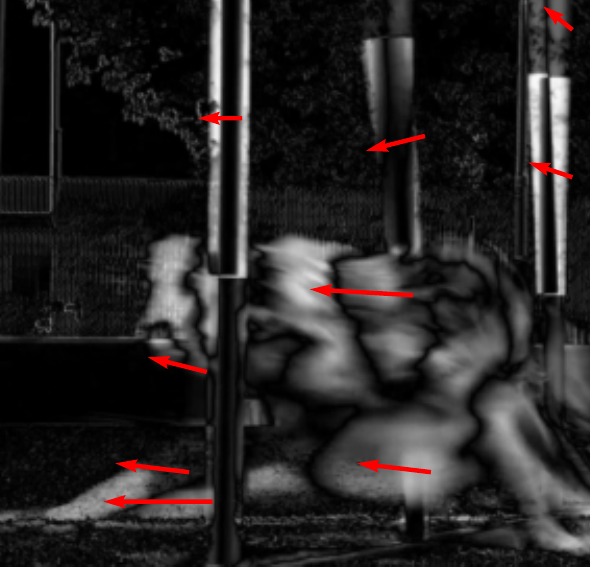}}
    \subfigure[17-th band]{
    \includegraphics[width=0.13\textwidth]{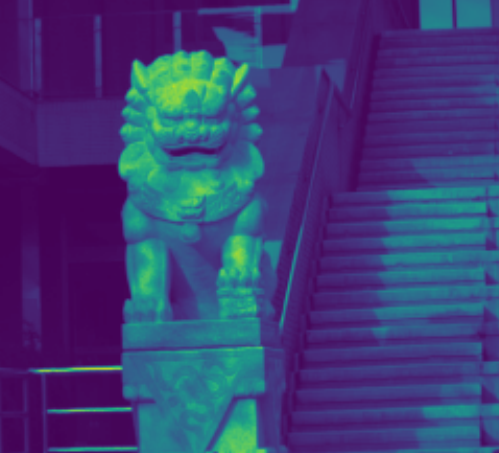}}
    \subfigure[18-th band]{
    \includegraphics[width=0.13\textwidth]{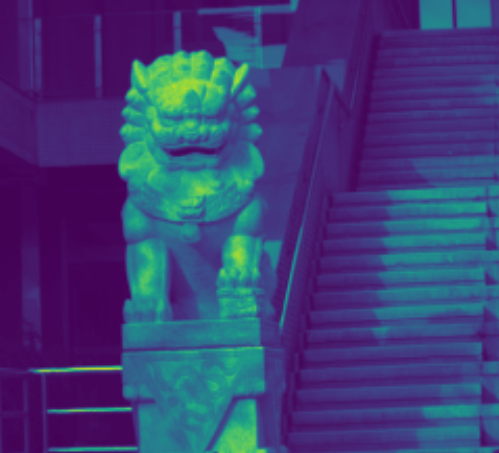}}
    \subfigure[diff. (reflectance)]{
    \includegraphics[width=0.13\textwidth]{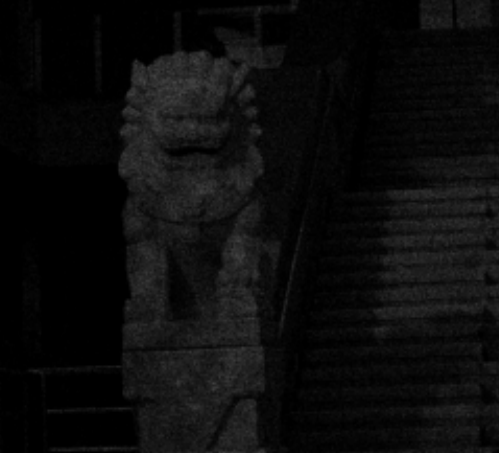}}
\caption{Exemplified differences of the video frames (up, dog) and HSI bands (bottom, lion). Temporal motion leads to the difference of video frames while the difference of HSI bands owes to reflectance variation without motion.}
\Description{}
\label{fig:frame diff}
\end{figure}

A practical HSI compression solution pursues 1) high-efficiency R-D performance and lightweight decoding complexity and 2) a negligible accuracy drop using decoded HSI for the downstream task. {In principle, each HSI collects a sequence of "frames" (spectral bands) at serial wavelengths, analogous to a video containing a sequence of frames at serial timestamps, which, however, differ fundamentally in inter-frame (spectral) correlations. Intuitively, the frame difference in a video mainly owes to the temporal motion, assuming consistent pixel intensity of objects across all timestamps. In contrast, such a "frame" difference in an HSI is due to reflectance variation at each pixel across spectral bands, typically assuming stationary scenes without temporal motion (see Fig.~\ref{fig:frame diff} and Fig.~\ref{fig:Spectral intensity}). Consequently, how to efficiently exploit correlations within and across spectral bands is crucial for improving compression performance and also benefiting downstream task on compressed HSI.}

{To this end, we propose {HINER}, a novel spectral-wise neural representation for HSI. The proposed {HINER} employs a positional encoding followed by an MLP to embed the spectral wavelengths of the input HSI sample explicitly. Such an explicit embedding is capable of effectively characterizing and exploiting cross-band correlation, which is then fed into a learnable neural decoder to generate the corresponding decoded HSI.}
{The pursuit of compression is achieved through collaboratively optimizing the encoder-decoder pair to generate more compact representations of spectral embedding and quantized decoder. Furthermore, we also propose to combine the {Content Angle Mapper} ({CAM}) measuring angle similarity between the reconstructed spectral band and its original counterpart and pixel-wise L1 loss, which contribute to maintaining global and local fidelity in signal reconstruction jointly.}

Simultaneously, the impairment of downstream task performance on lossy compressed HSIs is a practical challenge~\cite{guo2023hyperspectral}. Intuitively, the lossy compression may disrupt both the structural information and spectral continuity inherent in HSIs, which, without additional processing, will inevitably lead to accuracy degradation when optimized for vision tasks such as classification. To address this issue, we first employ a first-order Taylor expansion on the task accuracy degradation caused by the lossy compression, theoretically establishing an intrinsic connection between task accuracy and reconstruction fidelity. Then Adaptive Spectral Weighting ({ASW}) is introduced as the proxy of optimizing reconstructed HSIs with additional reconstruction loss as a optimized boundary constraint. Furthermore, building upon the monotonic mapping established by HINER between wavelengths and spectral bands, we propose {Implicit Spectral Interpolation} ({ISI}) as a data augmentation technique for training classification model on compressed HSIs. These result in significant enhancement in downstream performance.

% By introducing the {Adaptive Spectral Weighting} ({ASW}) as the proxy of optimizing reconstructed HSIs with additional reconstruction loss as a mathematical boundary constraint, the accuracy of downstream classification is greatly improved. Furthermore, building upon the monotonic mapping established by HINER between wavelengths and spectral bands, we propose {Implicit Spectral Interpolation} ({ISI}) as a data augmentation technique for training classification model on compressed HSIs, resulting in significant enhancement in overall accuracy.

{The main contributions of this paper are as follows:
\begin{enumerate}
    \item We propose {HINER}, a neural representation designed specifically for HSI. By introducing explicit encoding of wavelengths and global {CAM} loss, HINER effectively exploits spectral redundancy in HSI samples.
    \item We enhance the performance of downstream classification on lossy HSIs from two perspectives: adjusting the reconstruction to adapt to classification task through {ASW}, and improving the generalization of classification model with augmented data through {ISI}.
    \item Experimental results demonstrate the superior compression efficiency and comparable computational complexity of our proposed {HINER} compared to existing neural representation methods. Furthermore, there is a notable improvement in task accuracy when deploying classification on decoded HSIs at high compression ratios.
\end{enumerate}}

\begin{figure*}[htbp]
\centering
    \includegraphics[width=0.95\textwidth]{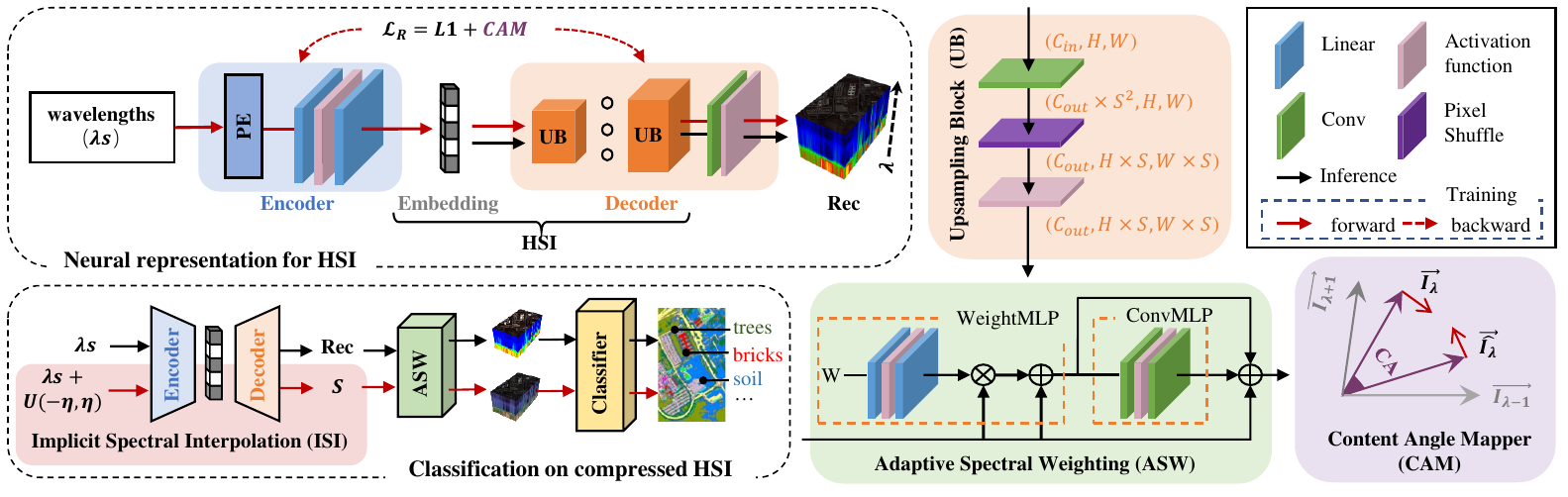}
\caption{The pipeline of our proposed {HINER}, the neural representation dedicated to compressing HSI, which also benefits downstream classification task on compressed HSI samples.}
\Description{}
\label{fig:pipeline}
\end{figure*}

\section{Related Work}
\subsection{Implicit Neural Representation}  \label{sec:related INR}
Implicit Neural Representations (INRs) have gained widespread interest for its remarkable capability in representing diverse multimedia signals, including
images~\cite{xie2023diner, liu2023finer}, videos~\cite{chen2021nerv, lee2023ffnerv, zhang2024boosting}, and neural radiance fields \cite{mildenhall2021nerf, wang2022nerf}. Among them, NeRV~\cite{chen2021nerv} proposed the first frame-wise INR for video, which took frame indices as inputs to generate corresponding RGB frames. Compared to previous pixel-wise INR methods (e.g., SIREN~\cite{sitzmann2020implicit}, Coin~\cite{dupont2021coin}), NeRV achieved better reconstruction quality while ensuring faster decoding. However, NeRV fully relied on the implicit learned decoder to characterize the input content and completely ignored the video content dynamics across frames. The subsequent HNeRV~\cite{chen2023hnerv} proposed to explicitly embed frame content instead of index, leading to better reconstruction and faster model convergence for video sequence. 
Some recent works also attempted to capture temporal correlation by frame difference~\cite{zhao2023dnerv}, optical flows~\cite{lee2023ffnerv}, etc. Recently, INR has also attracted practitioners in HSI, including super resolution~\cite{zhang2022implicit}, reconstruction~\cite{chen2023spectral}, fusion~\cite{wang2023ss, he2024two}, and compression~\cite{zhang2024compressing, rezasoltani2023hyperspectral, rezasoltani2023hyperspectral_v0}, showing remarkable potential in practical applications.

\subsection{HSI Compression} \label{sec:related compression}
HSI compression~\cite{rucker2005jpeg2000, meng2003efficient, penna2006progressive, christophe2008hyperspectral, qiao2014effective} commonly employs transform coding \cite{penna2007transform} to convert HSI in the pixel domain to a latent space (e.g., frequency domain).  Prominent transforms like Discrete Cosine Transform (DCT)~\cite{ahmed1974discrete} and Wavelet~\cite{farge1992wavelet} utilized linear transformations that were generally comprised of a set of linear and orthogonal bases. 
However, such a linear transformation with fixed bases might not fully exploit the redundancy because the content of the underlying image block was non-stationary and did not strictly adhere to the Gaussian distribution assumption~\cite{goyal2001theoretical}. Chakrabart et al.~\cite{chakrabarti2011statistics} and Guo et al.~\cite{guo2021learned} have demonstrated that real-world HSIs exhibited greater kurtosis and heavier tails than assumed Gaussian distribution, indicating a non-Gaussian nature of the HSI source.

Over the past few years, 
learning-based non-linear transform~\cite{dua2021convolution,palsson2018hyperspectral, la2022hyperspectral} have show potential in exploiting non-stationary content distribution.
Dua et al.~\cite{dua2021convolution} and La et al.~\cite{la2022hyperspectral} firstly introduced Auto-Encoder (AE) for lossy HSI compression. 
Subsequently, Variational Auto-Encoders (VAEs) incorporating variational Bayesian theory, turned to represent latent features of input HSIs from a probabilistic perspective. Building upon the VAE architecture, 
Guo et al.~\cite{guo2021learned} repurposed the hyperprior model~\cite{balle2018variational} to compress HSI, where the student's T distribution~\cite{yang2007student} replaced the original Gaussian distribution. Recently, Guo et al.~\cite{guo2023hyperspectral} further introduced contrastive learning to preserve spectral attributes.

\textbf{INR-based HSI compression.} INRs provide a novel perspective on HSI compression by translating it into model compression. For instance, aforementioned Zhang et al.~\cite{zhang2024compressing} and Rezasoltani et al.~\cite{rezasoltani2023hyperspectral, rezasoltani2023hyperspectral_v0} employed post-training quantization~\cite{ding2022towards, shi2023rate} to compress models. However, even 16-bit quantization still resulted in significant performance loss (sometimes exceeding 1dB), which also indicated quite limited model capability of pixel-wise INRs with fully MLP-based network architecture. Considering HSIs can be treated as sequences akin to videos, well-established INRs for video compression can also be applied in HSI compression (though sub-optimal, as will be discussed in Sec.~\ref{sec:preliminary}). Most video-based INRs follow a three-step compression pipeline: 1) pruning \cite{lu2022bayesian} to reduce model size; 2) quantization to reduce parameter bit-width; 3) entropy coding to reduce parameter statistical redundancy. Through these operations, the model is significantly compressed only with slight performance decline, attributed to elaborate network architecture enhancing model capability. For example, our proposed {HINER} subtly incorporates convolution, upsampling, GELU, etc., allowing the use of lower quantization bit-width (e.g., 8-bit) with a negligible reconstruction loss.

\subsection{HSI Classification} 
HSI classification, which assigns each spatial pixel to a specific class based on its spectral characteristics, is the most vibrant field of research in the hyperspectral community and has drawn widespread attention~\cite{ghamisi2017advances}. 
Extracting more discriminative features is recognized as a crucial procedure for HSI classification~\cite{li2019deep}, which achieves rapid advancements propelled by deep learning.

Many well-recognized networks have been widely and successfully applied in HSI classification, including CNN~\cite{yue2015spectral, slavkovikj2015hyperspectral, zhao2016spectral, zhong2017spectral, paoletti2018new}, AE~\cite{chen2014deep}, graph convolutional network (GCN)~\cite{hong2020graph}.
Recently, transformer-based methods~\cite{hong2021spectralformer, he2021spatial, wang2023dcn} show noticeable accuracy gains due to the self-attention mechanism, which effectively weights neighborhood information in dynamic input~\cite{lu2022high}. Hong et al.~\cite{hong2021spectralformer} developed a novel model called SpectralFormer (SF), capable of extracting features by aggregating multiple neighboring bands. 
% Additionally, SF implemented cross-layer skip connections to reduce information loss during layer-wise propagation. 
Given that SF currently exhibits leading performance, we employ it as our baseline classification model for downstream task evaluation.

\textbf{Classification on compressed HSI.} 
Most current approaches in HSI classification continue to rely on uncompressed data due to the observed accuracy degradation induced by lossy compression. Unlike RGB images which can be visually appreciated by humans, compressed HSI will become completely useless if it cannot be applied to downstream tasks. The idea of using compressed images for classification dates back to the last century~\cite{oehler1995combining}. While some studies have explored the impact of lossy compression on HSI classification outcomes~\cite{elkholy2019studying, zabala2011effects, wei2021effects, garcia2010impact, chen2019effects}, primarily focusing on predicting classification accuracy for a given compressed HSI, our emphasis is on mitigating degradation for specific compressed samples without uncompressed ground truth.

\section{Method} \label{sec:method}
{In this section, we begin by defining the optimization objective of neural representation for HSI (Sec.\ref{sec:preliminary}). Subsequently, {HINER}, a neural representation for HSI compression, is proposed by exploiting the correlations within and across spectral bands (Sec.~\ref{sec:HINER}). Lastly, We theoretically analyze and address the issue of accuracy degradation in classification task on compressed HSI. (Sec.~\ref{sec:compressive classification}). An overview of our overall pipeline is illustrated in Fig.~\ref{fig:pipeline}.}

\subsection{Preliminary} \label{sec:preliminary}

{Let $\boldsymbol{I} = \{I_{\lambda}\}_{\lambda=\alpha}^{\beta} \in \mathbb{R}^{H \times W \times C}$ denote an input HSI with spatial resolution of $H \times W$ and a total of $C$ spectral bands spanning the wavelength range $\lambda \in [\alpha, \beta]$. The objective of neural representation is to model a mapping function $\mathcal{F}$ from the embeddings $\boldsymbol{e}$ to the HSI $\boldsymbol{I}$: $\mathcal{F}(\boldsymbol{e}) \to \boldsymbol{I}$ using a neural network. This work suggests to process HSI spectral-wisely.}
Given the spectral band $I_\lambda$ with wavelength $\lambda$, a learnable decoder $\mathcal{D}(\cdot)$ is employed for reconstruction by inputting spectral embedding $e_\lambda$. Our goal is to minimize the distortion between the input $I_\lambda$ and its reconstructed counterpart $\hat{I}_\lambda$ with the restricted model parameter $\theta$. As a result, the rate-distortion (R-D) optimization objective can be formulated as:
\begin{align}
    \mathop{\arg\min} \sum_{\lambda = \alpha}^{\beta} & \mathcal{L}(I_\lambda, \hat{I}_\lambda) 
    = \mathop{\arg\min}_{\boldsymbol{e}, \mathcal{D}} \sum_{\lambda = \alpha}^{\beta} \mathcal{L} \left(I_\lambda, \mathcal{D} (e_\lambda)\right),  \nonumber \\
    & \operatorname{ s.t. } \ \ \ \theta(\mathcal{\boldsymbol{e}}) + \theta(\mathcal{D}) \leqslant \theta, \label{eq: RD objective}
\end{align}
where $\mathcal{L}$ represents the distortion loss. The bitrate $\theta(\mathcal{\boldsymbol{e}})$ used for embeddings and the decoder parameter $\theta(\mathcal{D})$ collectively comprise the overall bitrate consumption, subject to the constraint of $\theta$. 

As a comparative neural representation in video compression, NeRV~\cite{chen2021nerv} completely relied on a learnable decoder for implicit representation without any content embeddings. Since the embedding is generated by fixed position encoding of temporal indices, {the only consumed bitrate is $\theta(\mathcal{D})$ without $\theta(\mathcal{\boldsymbol{e}})$. HNeRV~\cite{chen2023hnerv} firstly proposed the hybrid neural representation framework, which incorporated a learnable encoder to produce additional embeddings from frame content. Through a small amount of bitrate consumption by $\theta(\mathcal{\boldsymbol{e}})$, such explicit content embeddings greatly improved the coding efficiency and model convergence.}

\begin{figure}[tbp]
\centering
\subfigure[Video]{
\includegraphics[scale=0.29]{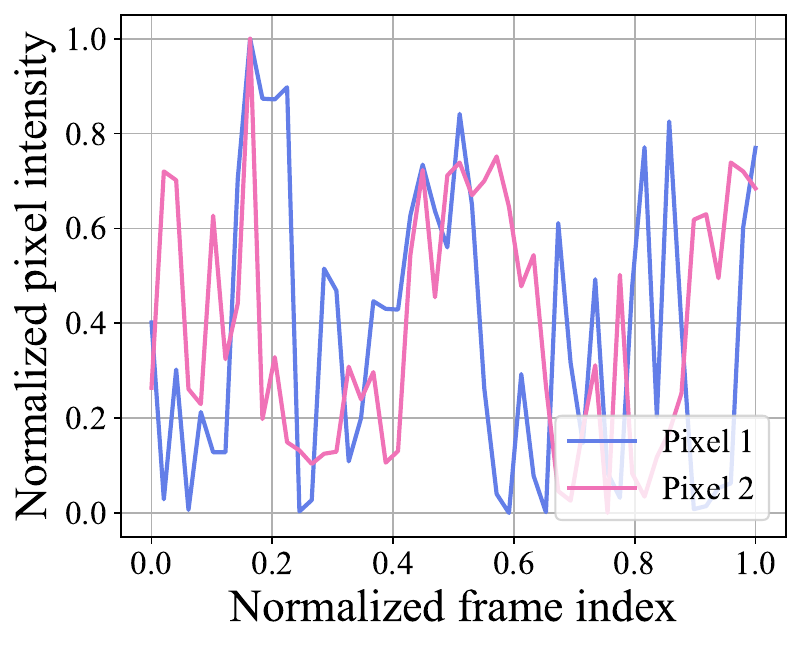}\label{fig:Spectral intensity a}}
\subfigure[HSI]{
\includegraphics[scale=0.29]{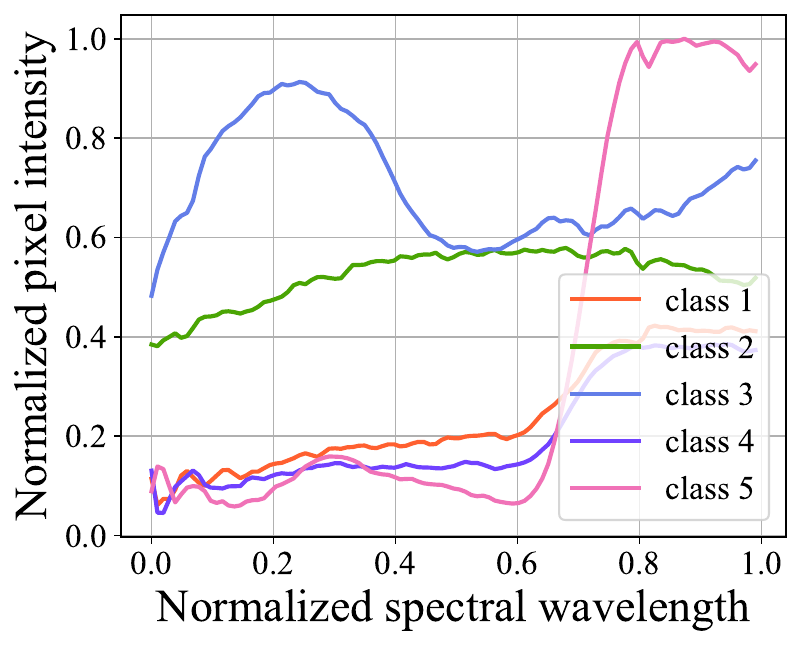}\label{fig:Spectral intensity b}}
\caption{{Pixel intensity distribution in fixed spatial position.}}
\label{fig:Spectral intensity}
\Description{}
\end{figure}

{Although explicit content embedding has demonstrated remarkable performance in video compression, an accompanying issue has arisen: can this success be replicated directly on HSI?
As mentioned above, frame differences in a video primarily stem from non-monotonic temporal motion, which makes monotonic frame indices inadequate for capturing pixel correlations among neighboring frames (see Fig.~\ref{fig:Spectral intensity a}). Therefore, capturing differentiated content from each frame can yield better temporal embedding compared to content-agnostic frame indices. Conversely, in HSI, such "frame" differences originate from reflectance variation at each object across spectral wavelengths, where we often assume stationary objects without temporal motion. As can be seen in Fig.~\ref{fig:Spectral intensity b}, there is a potential mapping relationship between wavelengths and pixel intensities (each pixel in HSI corresponds to a specific object class, such as tree, soil, etc.). Consequently, content embedding is sub-optimal for representing HSI which requires fully leveraging spectral correlation. To address this, we propose {HINER}, a neural representation fully exploiting spectral redundancy in Sec.~\ref{sec:HINER}. Furthermore, we theoretically investigate and overcome the problem of performance degradation in downstream classification on compressed HSIs built upon the characteristic of HINER in Sec.~\ref{sec:compressive classification}.}

\subsection{Neural Representation for HSI\label{sec:HINER}}

\textbf{Spectral Wavelength Embedding.}
To capture spectral correlation, we take a straightforward yet highly effective approach by \emph{explicitly embedding spectral wavelength $\lambda$}.

In the specific implementation, {HINER} explicitly encodes the normalized $\lambda \sim U(0,1)$ using a learnable encoder $\mathcal{E}$ to generate the spectral embeddings $\boldsymbol{e} = \{e_\lambda\}_{\lambda=\alpha}^{\beta}$, which is then forwarded to the decoder $\mathcal{D}$ for the reconstruction of the HSI. {According to Eq.~\eqref{eq: RD objective}, the encoder does not consume bitrate and is only used to produce the spectral embeddings that need to be further encoded. However, considering that the training time of the entire neural representation model is equivalent to the encoding time of the HSI, it is crucial to design an efficient encoder with the following two characteristics: \emph{1) Maintaining a low level of computational complexity;} \emph{2) Efficiently capturing spectral correlation.}}

Inspired by the practice of \cite{mildenhall2021nerf} in neural radiance fields, $\mathcal{E}$ is built as a composition of two functions $\mathcal{E} = \mathcal{M} \circ \mathcal{P}$, by which
\begin{align}
    e_\lambda = \mathcal{E}(\lambda) =  \mathcal{M} \left(\mathcal{P}(\lambda)\right),
\end{align}
where $\mathcal{M}$ stands for a tiny learnable MLP layer, and $\mathcal{P}$ denotes the frequency Positional Encoding (PE) \cite{mildenhall2021nerf, tancik2020fourier} to map $\lambda$ into a higher dimensional space $\mathcal{P}: \mathbb{R} \to \mathbb{R}^{2l}$, i.e.,
\begin{align}
    \mathcal{P}(\lambda) \!=\! \left( \sin(b^0\pi \lambda), \! \cos(b^0\pi \lambda),\right.\!\left.\!\dots, \! \sin(b^{l-1}\pi \lambda), \! \cos(b^{l-1}\pi \lambda) \right)\!.
\end{align}
{The rationale behind not directly inputting $\lambda$s into the MLP layer without positional encoding is due to the well-known spectral bias~\cite{rahaman2019spectral, xie2023diner} in MLP. This bias tends to prioritize learning low-frequency components of the signal, potentially leading to the network's inability to adequately model high-frequency variation.~\cite{rahaman2019spectral, mildenhall2021nerf, tancik2020fourier}. This phenomenon is detailed in supplementary material.}

By jointly optimizing the encoder and decoder, such a lightweight encoder is sufficient for information extraction and facilitates faster model convergence in training, which is also known as the encoding process for INR methods. The bitrate overhead of such spectral embedding is negligible (see Sec.~\ref{sec:3.3}), but it dramatically improves the performance of {HINER} by exploiting the inter-band correlation for better coding efficiency.

\textbf{Content Angle Mapper.} 
In general, INR models are typically optimized using L-p loss function~\cite{li2020norm, chen2023hnerv, rezasoltani2023hyperspectral_v0}. {However, the pixel-wise L-p loss lacks the ability to supervise global content similarity within spectral band. To address this limitation, some recent works~\cite{nilsson2020understanding, chen2021nerv} introduce the Structure Similarity Index Measure (SSIM) loss by considering the correlations in luminance, contrast, and structure of the images, which may not apply to single spectral band. Drawing inspiration from the spectral angle mapper used for pixel correlation analysis~\cite{wang2024enhancing, wang2021ssconv}, we introduce the {Content Angle Mapper (CAM)} to calculate the angle between the original spectral band vector $\Vec{I}_\lambda$ and its reconstructed counterpart $\Vec{\hat{I}}_\lambda$. Minimizing {CAM} enables us to spatially exploit global content correlation in each spectral band. Additionally, L1 loss is also incorporated for pixel-wise supervision, which is proved to be more appropriate for scenes characterized by complex textures with high-frequency information~\cite{zhao2023dnerv, candes2005decoding}.} 
Consequently, the optimization objective for training {HINER} can be formulated as:
\begin{align}
    \mathcal{L}_{R} = \underbrace{\sum_{\lambda=\alpha}^{\beta} || \hat{I}_\lambda - I_\lambda ||}_{L1 \ \ loss} 
    + \gamma \cdot
    \underbrace{\sum_{\lambda=\alpha}^{\beta} \frac{180}{\pi} \arccos \left( \frac{\Vec{\hat{I}}_\lambda^T \cdot \Vec{I}_\lambda}{\Vert \Vec{\hat{I}}_\lambda^T \Vert_2 \Vert \Vec{I}_\lambda \Vert_2}  \right)}_{CAM\ \ },
\end{align}
where the vector $\Vec{I}_\lambda \in \mathbb{R}^{m \times 1}$ denotes the flattened spectral band $I_\lambda$, $m = H \times W$ is determined by spatial resolution, and $\gamma$ is introduced to make a trade-off between these two losses.

\textbf{Compression.} 
To further reduce the actual bitrate consumption of our {HINER}, we follow HNeRV and employ the same quantization and entropy coding methods for model compression.
In model quantization, the floating-point vector $\boldsymbol{\mu}_{float}$ (e.g., weight or bias in a convolutional layer) is quantized using:
\begin{align}
    \boldsymbol{\mu}_{int} &=  clip \left( \lfloor \frac{{\boldsymbol{\mu}}_{float} - \min(\boldsymbol{\mu})}{s_{\boldsymbol{\mu}}} \rceil , 0, 2^b - 1 \right), \nonumber \\
    ~
    &\operatorname{ where } \ \ \ s_{\boldsymbol{\mu}} = \frac{\max(\boldsymbol{\mu}) - \min(\boldsymbol{\mu})}{2^b - 1},
\end{align}
$\lfloor \cdot \rceil$ rounds the input to the nearest integer. $b$ denotes the quantization bit-width, and $s_{\boldsymbol{\mu}}$ is the linear scaling factor. We utilize the Huffman~\cite{huffman1952method} coding as the lossless entropy coding method to further compress model parameters after quantization.

\subsection{Classification on Compressed HSI} \label{sec:compressive classification}

HSI classification refers to assigning a predefined label to each individual pixel~\cite{fauvel2012advances}, which is similar to the semantic segmentation task for the RGB image. The degradation in classification performance on compressed HSIs is related to the intrinsic characteristics of HSI, where pixel intensity corresponds to the spectral reflectance of objects across multiple spectral bands \cite{guo2023hyperspectral}. As depicted in Fig.~\ref{fig:Spectral intensity}, two categories with similar spectral reflectance (e.g., class 1 and class 4) may become indistinguishable after lossy compression. We enhance the performance of classification on lossy HSIs from two perspectives: 1) adjusting the compressed reconstruction to adapt to classification task; 2) improving the generalization of classification model with augmented data.

\textbf{Adaptive Spectral Weighting (ASW).} 
{We first analyse classification loss $\mathcal{L}_C$ theoretically. Having the classification model parameterized by $\boldsymbol{\theta}$, the original uncompressed HSI $\boldsymbol{I}$, and the reconstructed HSI $\hat{\boldsymbol{I}}$, we employ additive noise~\cite{zhang2021rethinking, wei2021qdrop} to model the compression loss, i.e., $\boldsymbol{I} = \hat{\boldsymbol{I}} + \boldsymbol{u}(\hat{\boldsymbol{I}})$. Consequently, the task performance degradation induced by compression can be defined as:}
\begin{align}
    \mathbb{E} \left[ \mathcal{L}_C(\boldsymbol{\theta}, \boldsymbol{I}) -\mathcal{L}_C(\boldsymbol{\theta}, \hat{\boldsymbol{I}}) \right].
    \label{eq:lossy compression}
\end{align}
With the first-order expansion on Eq.~\eqref{eq:lossy compression}, we will derive:
\begin{align}
    \mathcal{L}_C(\boldsymbol{\theta}, \boldsymbol{I}) -\mathcal{L}_C(\boldsymbol{\theta}, \hat{\boldsymbol{I}})
    \approx
    \nabla_{\hat{\boldsymbol{I}}} \mathcal{L}_C(\boldsymbol{\theta}, \hat{\boldsymbol{I}})^T \cdot \boldsymbol{u}(\hat{\boldsymbol{I}}),
    \label{eq:expansion}
\end{align}
An intuitive solution is to optimize the input $\hat{\boldsymbol{I}}$ based on the gradient $\nabla_{\hat{\boldsymbol{I}}} \mathcal{L}_C(\boldsymbol{\theta}, \hat{\boldsymbol{I}})^T$ while ensuring $\hat{\boldsymbol{I}}$ close to $\boldsymbol{I}$, characterized by $||\boldsymbol{I}-\hat{\boldsymbol{I}}|| = ||\boldsymbol{u}(\hat{\boldsymbol{I}})||<\epsilon$. However, directly optimizing $\hat{\boldsymbol{I}}$ can be challenging to converge due to its high spatial and spectral resolution.
Thus, we introduce a tiny learnable {Adaptive Spectral Weighting} ({ASW}) module to adjust $\hat{\boldsymbol{I}}$, as shown in Fig.~\ref{fig:pipeline}. Then, the optimization of $\hat{\boldsymbol{I}}$ is converted into the optimization of network parameters, which can be easily solved by gradient descent, where optimizing $\mathcal{L}_C$ aims to make $\nabla\mathcal{L}_C \to 0$ and optimizing $\mathcal{L}_{R}$ aims to constrain $\boldsymbol{u}$:
\begin{align}
    \theta_{ASW} = \mathop{\arg\min} \mathcal{L}_{C} + \beta \cdot \mathcal{L}_{R},
    \label{eq: awn}
\end{align}

{ASW} first spectral-wisely re-weights the reconstructed HSI by multiplying learned factors, followed by an MLP comprising 1x1 conv for cross-spectral information aggregation. The rationale behind {ASW} lies in the varying importance of spectral bands for reconstruction and downstream classification~\cite{zhou2023btc}. Thus, ASW facilitates the translation from perception-oriented reconstruction to classification-oriented reconstruction. More details can be found in the supplementary material.

\begin{figure}[tbp]
\centering
\subfigure[Trained using reconstructed HSI]{
\label{fig:w/o isi}
\includegraphics[width=0.2\textwidth]{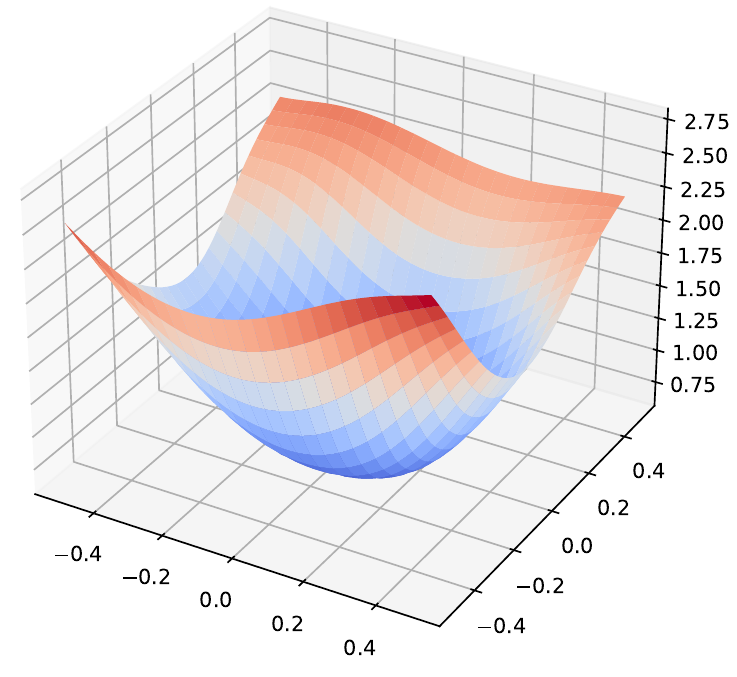}}
\subfigure[Trained w/ {ISI} (flatter)]{
\label{fig:w/ isi}
\includegraphics[width=0.2\textwidth]{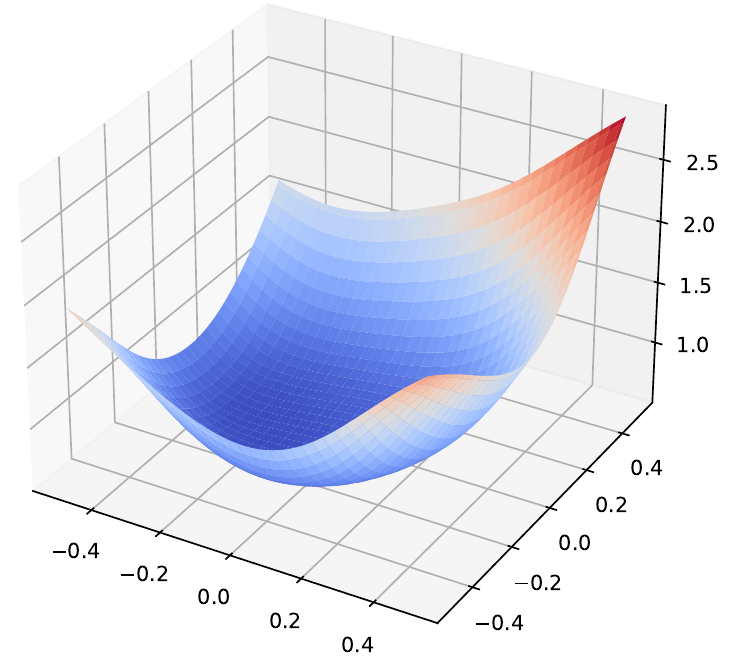}}
\caption{\textbf{Loss landscape~\cite{li2018visualizing} of trained classification model.}}
\Description{}
\label{fig:loss surface}
\end{figure}

\begin{figure*}[t]
\centering
\subfigure[Indian Pines dataset]{
\label{fig:compress indian}
\includegraphics[scale=0.4]{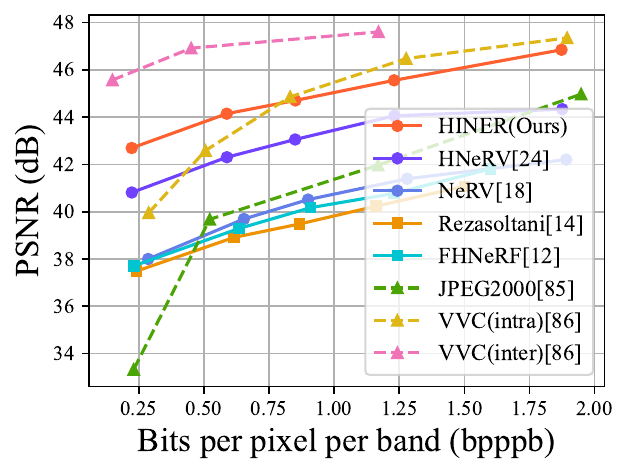}}
\subfigure[Pavia University dataset]{
\label{fig:compress paviau}
\includegraphics[scale=0.4]{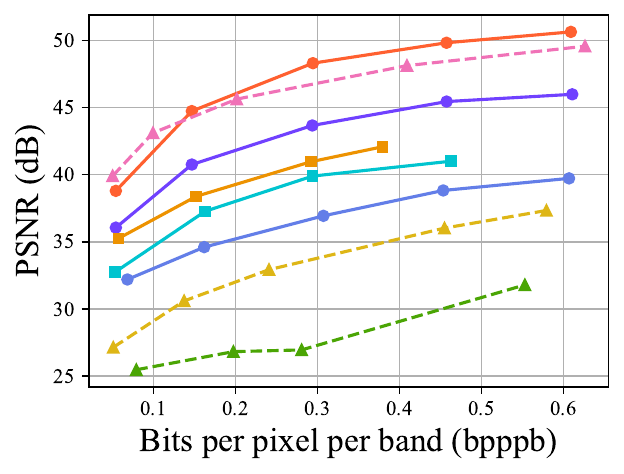}}
\subfigure[Pavia Centre dataset]{
\label{fig:compress pavia}
\includegraphics[scale=0.4]{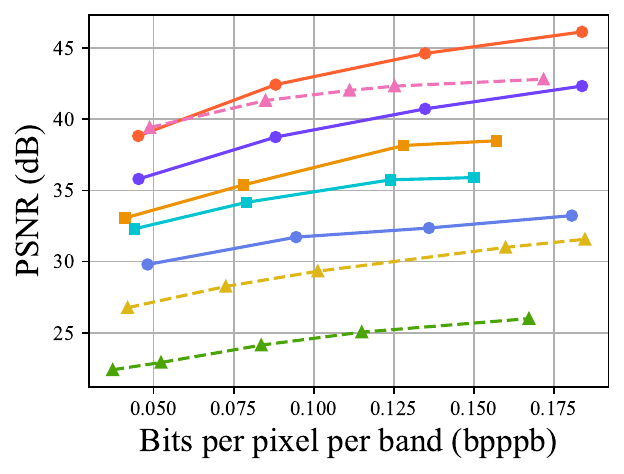}}
\subfigure[CHILD dataset]{
\label{fig:compress child}
\includegraphics[scale=0.4]{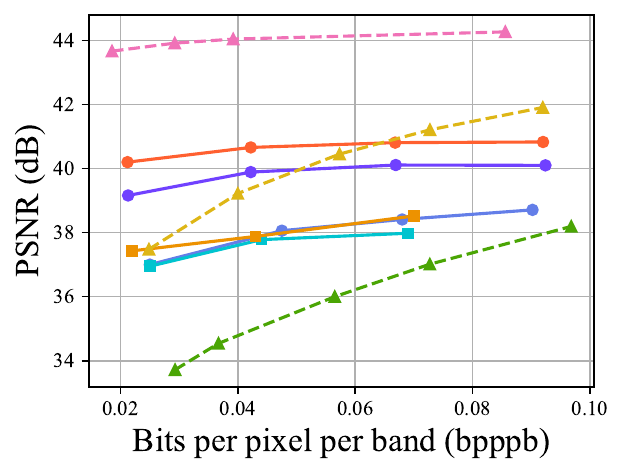}}
\caption{R-D performance comparisons across HSI datasets.}
\Description{}
\label{fig:compress result}
\end{figure*}

\begin{figure*}[htbp]
\centering
    \includegraphics[width=0.95\textwidth]{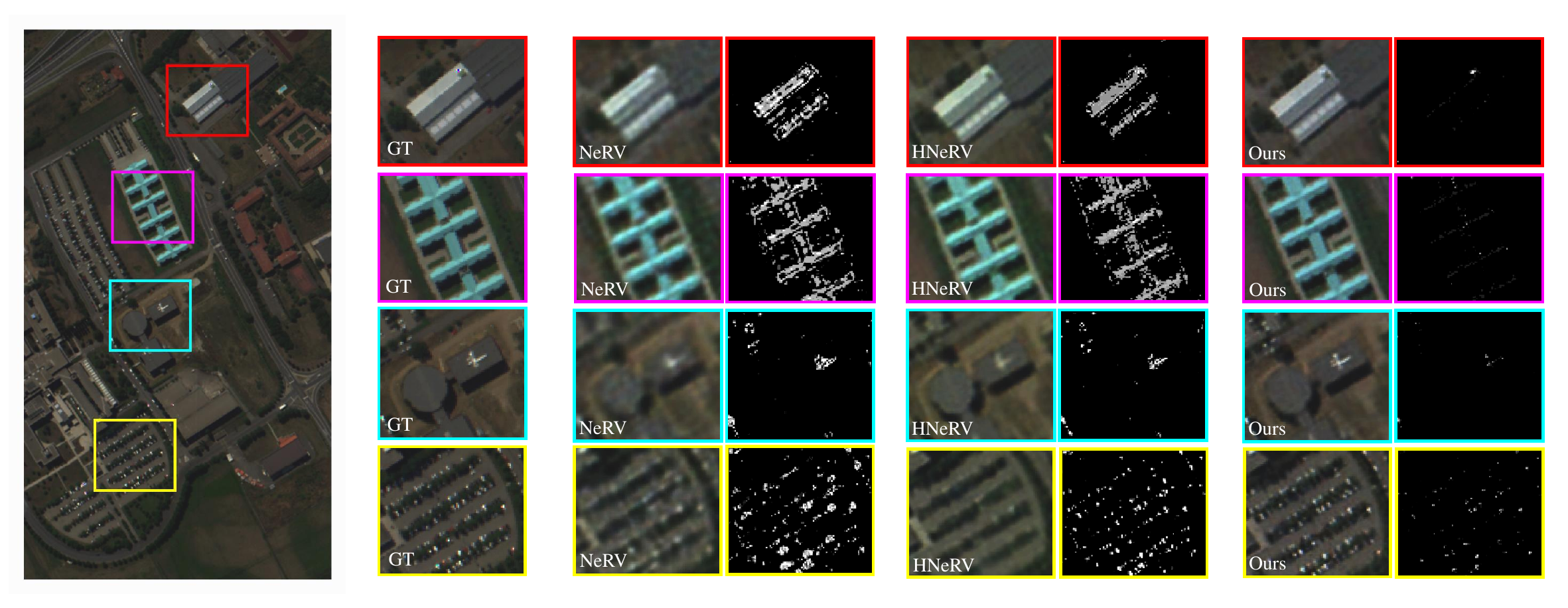}
\caption{Visualization comparisons of NeRV, HNeRV and our {HINER} on Pavia University dataset. The residual between the reconstruction and the ground truth (GT) is accompanied by each reconstructed result.}
\Description{}
\label{fig:visualization}
\end{figure*}

\textbf{Implicit Spectral Interpolation (ISI).} %Implicit Spectral Augmentation
{HINER} establishes a monotonic continuous mapping from spetral wavelengths to spectral bands. This enables {HINER} to reconstruct corresponding spectral bands for arbitrary wavelengths, even if these wavelengths or bands do not exist in the original discrete HSI sample (in some literatures, this function is also refered to as reconstruction~\cite{robles2015single} or super-resolution \cite{wang2023learning}). Leveraging this continuous mapping, we can construct an augmented sample set $\mathcal{S}$ containing multiple randomly sampled HSI by adding random variables to the input wavelengths
\begin{align}
    \mathcal{S} = \{\mathcal{HINER} \left( \lambda + U(-\eta, \eta)\right) | \lambda \in (0, 1)\},
\end{align}
where $U(-\eta, \eta)$ represent a uniform distribution. Training the classification model with the augmented data set $\mathcal{S}$ significantly enhances performance on compressed HSI. It's important to note that we do not introduce ground truth HSI during training, thus improving the practical applicability of {ISI}. The rationale behind {ISI} lies in the fact that data augmentation can enhance the model's generalization on compressed HSI data~\cite{neyshabur2017exploring, jiang2019fantastic, wei2021qdrop, wang2022semantic}, thereby leading to improved accuracy. One intuitive manifestation of generalization is the flatness of the loss landscape~\cite{li2018visualizing}. A flatter loss landscape, indicative of better generalization, exhibits relatively small loss changes under parameter perturbations, whereas a sharp loss landscape indicates otherwise. As depicted in Fig.~\ref{fig:loss surface}, the classification model trained with {ISI} exhibits a flatter loss landscape, in which Fig.~\ref{fig:w/o isi} can also be considered a special case of $\eta=0$. We provide detailed explanation in the supplementary material.

\subsection{Discussion}

{So far, we have developed a neural representation framework, {HINER}, with a series of optimizations tailored for HSI, aligning better with its intrinsic characteristics and also aiding in maintaining performance of downstream tasks on compressed reconstruction. Traditional pixel-wise INRs~\cite{zhang2024compressing,rezasoltani2023hyperspectral_v0} and content-embedded HNeRV~\cite{chen2023hnerv} neglect inter-spectral correlations, while content- independent NeRV~\cite{chen2021nerv} can not activate the decoder well for characterization of different spectrum using position encoding. HINER attempt to explicitly exploit cross-spectral correlations through wavelength embedding and fully supervise local and global reconstruction fidelity within a specific band by combining L1 and {CAM} loss. We also thoroughly consider the degradation of downstream classification task caused by lossy compression. By using {ASW} for reweighting the compressed reconstruction and {ISI} for data augmentation in training classification model, we effectively address the issue of loss in task accuracy, which was not mentioned in previous works~\cite{zhang2024compressing,rezasoltani2023hyperspectral_v0}.}
For a fair comparison, we adopted the same bit-rate controlling strategy as
HNeRV: keep fixed 8-bit parameter bit-width and vary the parameter quantity to control the rate, where we quantize both the model and the embedding into 8 bits. 
% Balancing bit-width and parameter quantity is a valuable area for future research.

\section{Experiments}
\label{sec:Experiments}

\subsection{Setup}
\textbf{Datasets.} We conduct the evaluation on four popular HSI datasets with varying  resolutions: 1) The {Indian Pines} dataset with size of $145\times 145\times200$; 2) The \emph{Pavia University} dataset with size of $610 \times 340\times103$; 3) The \emph{Pavia Centre} dataset with size of $1096 \times 715\times102$; 4) The \emph{CHILD} \cite{huang2022high} dataset with size of $960\times 1056\times145$. 

\textbf{Metrics.} For compression performance comparison, \emph{Peak Signal-to-Noise Ratio (PSNR)} is used to measure the reconstruction quality, and \emph{bits per pixel per band (bpppb)} reports the consumption of compressed bitrate.  The classification performance on compressed HSI is evaluated using \emph{Overall Accuracy (OA)}, \emph{Average Accuracy (AA)}, and \emph{Kappa Coefficient ($\kappa$)}. 

\textbf{Baselines.} For compression performance comparison, we select the frame-wise neural representation methods, namely NeRV~\cite{chen2021nerv} and HNeRV~\cite{chen2023hnerv}, as well as the pixel-wise methods specifically designed for HSI, FHNeRF~\cite{zhang2024compressing} and Rezasoltani~\cite{rezasoltani2023hyperspectral_v0}.  JPEG2000~\cite{du2007hyperspectral} and VVC~\cite{bross2021overview} are exemplified to represent traditional image and video codecs. 
For JPEG2000, we employ OpenJPEG to independently compress each spectral band. For VVC, we utilize the reference software VTM 16.0 and conduct compression experiments with both intra and inter (i.e., Random Access) profiles. 
% The intra profile compresses each spectral band separately while the inter profile processes HSIs in spectral order as the coding of videos. 
For downstream classification, we use SpectralFormer (SF)~\cite{hong2021spectralformer} as the baseline, known for its leading performance.

\textbf{Implementation.} 
We faithfully reproduce the compared methods following their default settings on the HSI dataset. For the training of {HINER}, we adopt the Adam optimizer~\cite{kingma2014adam} with a batch size of 1. The initial learning rate is 0.001 with a cosine descent strategy. Stride sizes used for upsampling in the decoder are configured at (5, 3, 2, 2) for Indian Pines, (5, 4, 3, 2) for Pavia University, (5, 4, 3, 2, 2) for Pavia Centre, and (5, 4, 4, 2, 2) for CHILD, respectively. Unless specified otherwise, all experiments are conducted using PyTorch with an Nvidia RTX 3090 for totally 300 epochs. For classification, the learning rate is 0.0005 with 0.005 weight decay. Epochs for Indain Pines and Pavia University are 300 and 480, respectively.

\begin{figure}[tbp]
\centering
\subfigure[Encoding Time]{
\label{fig:Encoding Time}
\includegraphics[scale=0.3]{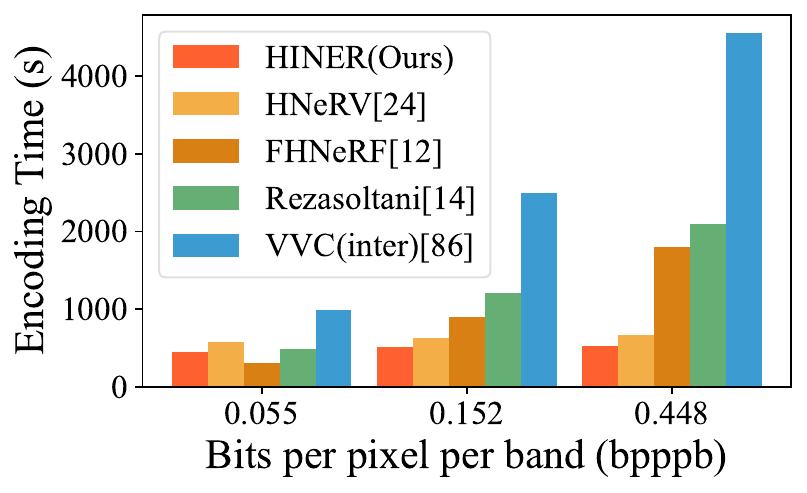}}
\subfigure[Decoding Time]{
\label{fig:Decoding Time}
\includegraphics[scale=0.3]{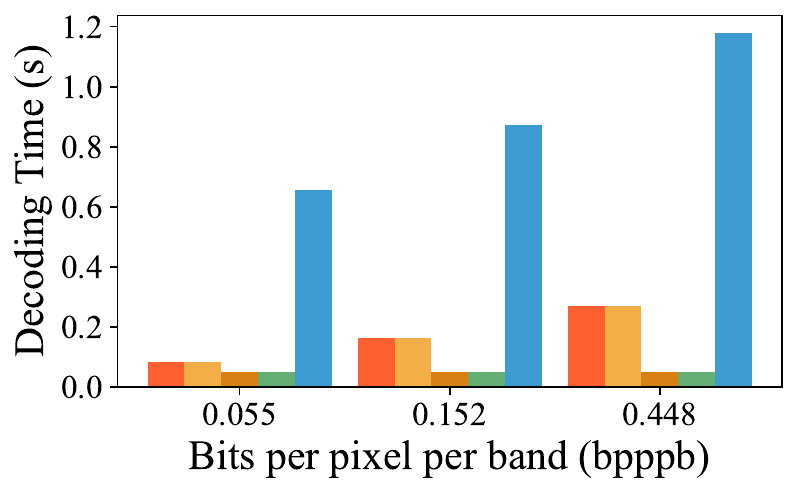}}
\caption{{Encoding \& decoding complexity comparisons.}}
\Description{}
\label{fig:E and D time}
\end{figure}

\subsection{HSI Compression}  \label{sec:experiment compression}
\textbf{Performance.} We present the R-D curves in Fig.~\ref{fig:compress result} across various datasets. The proposed {HINER} clearly outperforms other neural representation methods. Leveraging efficient spectral wavelength embeddings, {HINER} not only significantly surpasses the content-agnostic method (i.e., NeRV), but also proves better than the content embedding method (i.e., HNeRV) in the HSI dataset. Compared with pixel-wise FHNeRF and Rezasoltani, such a band-wise representation of {HINER} exhibits superior rate-distortion advantage. Furthermore, our method is far better than the earlier image codec JPEG2000 across all datasets and superior to the VVC intra coding in Pavia University and Pavia Centre, in which our {HINER} is even comparable with the VVC inter coding. However, there is still a performance gap between learning-based methods and the VVC inter coding in Indian Pines and CHILD. One possible reason is that these two datasets capture simple scenarios with few texture information, which is easy for {motion prediction} in VVC thus greatly improving the coding efficiency. Fig.~\ref{fig:visualization} visualizes the reconstruction results of neural representation methods on Pavia University. Notably, our method exhibits a much closer reconstruction to the original data.

\textbf{Complexity.} We also report the encoding and decoding time of our {HINER}, as well as HNeRV, FNeRF, Rezasoltani, and VVC inter profile in Fig.~\ref{fig:E and D time}. Our model presents faster encoding and decoding compared to VVC, e.g., up to $10\times/6\times$ encoding/decoding time reduction. The encoding of ours is also faster than HNeRV due to the lightweight spectral encoder. When compared with FHNeRF and Rezasoltani, spectral-wise {HINER} exhibits faster encoding, and the gap further increases with model size. Although the inclusion of the upsampling and GELU~\cite{hendrycks2016gaussian} operation slows down our decoding than fully MLP-based FHNeRF and Rezasoltani, {HINER} achieves a PSNR improvement of more than 5 dB while still maintaining over 400 band-per-second decoding speed at Pavia University.

\begin{figure}
    \centering
    \includegraphics[scale=0.5]{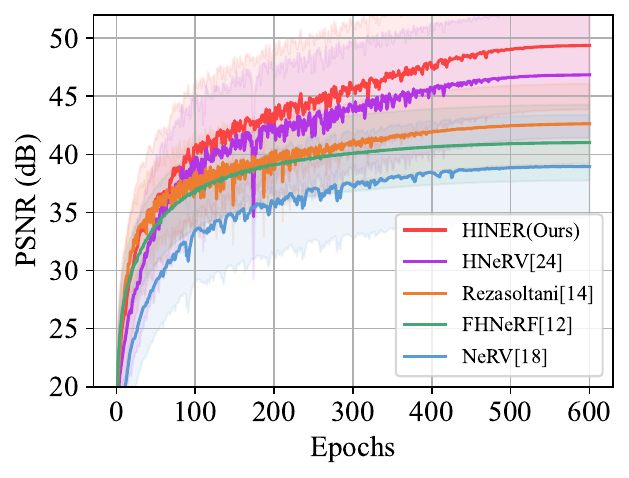}
    \caption{Regression capacity comparisons.}
    \label{fig:regression}
    \Description{}
\end{figure}

\textbf{Regression.}
To evaluate the efficiency of {HINER}, we conduct comparisons regarding regression capacity in Fig.~\ref{fig:regression}. As shown, those methods incorporating additional information embeddings, i.e., HNeRV and ours, lead to better reconstruction quality and faster model convergence speed than others. Moreover, our method shows the best performance, indicating the effectiveness of spectral embedding for HSI representation. An interesting observation is that Rezasoltani and FHNeRF exhibit rapid saturation in earlier training, probably due to pure MLP architecture limits the model capacity, and such pixel-wise representation is inadequate to capture cross-spectral redundancies.

\subsection{Classification on Compressed HSI}
% When lossy compressed HSIs are widely used for storage and transmission due to their lower space-time resource occupation, downstream servers will not access the original ground truth data. Thus, related services, such as classification, have to rely solely on the compressed HSI. 
As shown in Table~\ref{tab:classification}, lossy compression results in a significant degradation when optimized for classification (SF$^\clubsuit$ vs. SF), primarily due to the spectral information that determines the class of objects has been compromised. Our proposed ASW and ISI, effectively alleviate degradation and maintain considerable accuracy even under high compression ratios, e.g., up to $\times 109$, approaching the levels trained with ground truth. Notably, our approach even surpasses the performance of SF trained with ground truth data in the IndianPine. One potential explanation for this phenomenon is that our theoretical framework aids in mitigating data bias~\cite{jiang2020identifying, Kim_2019_CVPR} between the training and testing sets to some extent, when $\boldsymbol{u}(\boldsymbol{I})$ in Eq.~\ref{eq:expansion} is interpreted as a measure of data bias. This is an interesting topic for further study in the future.

\begin{table}[htbp]
%\footnotesize
  \centering
  \caption{Quantitative performance of classification. $\clubsuit$ represents the results trained with compressed HSI without ground truth. CR denotes compression ratio. 
  % The best results are highlighted in \sethlcolor{pink}\hl{bold}.
  }
    \begin{tabular}{clcccc}
    \toprule
    \multicolumn{1}{l}{Datasets} & Methods & CR & OA (\%)    & AA (\%)    & $\kappa$ \\
    \midrule
    \midrule
    \multirow{2}[6]{*}{IndianP} & SF    & $\times 1$ & 81.76 & 87.81 & 0.7919 \\
\cmidrule{2-6}          & SF $^\clubsuit$  & \multirow{2}[2]{*}{$\times 28$} & 79.15 & 84.27 & 0.7633 \\
          & Ours $^\clubsuit$  &       & \cellcolor{pink}\textbf{87.03}  & \cellcolor{pink}\textbf{90.99} & \cellcolor{pink}\textbf{0.8519} \\
% \cmidrule{2-6}          & SF $\star$   & \multirow{2}[2]{*}{} &       &       &  \\
%           & Ours $\star$  &       &       &       &  \\
    \midrule
    \multirow{3}[4]{*}{PaviaU} & SF    & $\times 1$ & 91.07 & 90.20  & 0.8805 \\
\cmidrule{2-6}          & SF $^\clubsuit$   & \multirow{2}[2]{*}{$\times 109$} & 86.29 & 87.89 & 0.8203 \\
          & Ours $^\clubsuit$  &       & \cellcolor{pink}\textbf{88.93}  & \cellcolor{pink}\textbf{88.96} & \cellcolor{pink}\textbf{0.8529}  \\
    \bottomrule
    \end{tabular}%
  \label{tab:classification}%
\end{table}%

\subsection{Ablation Study \label{sec:3.3}}

\textbf{Spectral Wavelength Embedding.}
As mentioned above, by explicitly encoding the spectral wavelength $\lambda$s, the spectral correlation is embedded to assist the decoder reconstruction. One deduction is that when we randomly reorder the spectral bands, i.e., shuffling the original mapping from wavelengths to corresponding bands, the permutation of spectra would be disturbed, thereby affecting the reconstruction results of {HINER}. 
The \emph{case 1} in Table~\ref{tab:compress ablation} confirms this deduction, in which our performance suffers from the shuffle operation compared to the default configuration. However, as a comparison, HNeRV is immune from the band shuffle without any performance loss, indicating that content embeddings fail to capture the inter-band correlation.

In addition, we also examine the effectiveness of the explicit encoder by \emph{case 2} in Table~\ref{tab:compress ablation}. We solely remove the encoder $\mathcal{E}$ so that the input HSI is fully represented by the decoder with the fixed position encoding as in NeRV. As observed, such a pattern greatly decreases the coding efficiency, which illustrates the effectiveness of our explicit encoder in learning spectral wavelength embeddings.

\begin{table}[htbp]
%\footnotesize
  \centering
    \caption{Ablations on spectral wavelength embedding.}
    \begin{tabular}{cccccc}
    \toprule
          &  & w/o shuffle & w/ $\mathcal{E}$  & Indian & PaviaU \\
    \midrule
    \midrule
    \multirow{2}[2]{*}{HNeRV} & case 1 & \XSolidBrush & \Checkmark & 44.33 & 43.66 \\
          & default & \Checkmark & \Checkmark & 44.33 & 43.66 \\
    \midrule
    \midrule
    \multirow{3}[2]{*}{{HINER}} & case 1 & \XSolidBrush & \Checkmark & 45.50 & 46.52 \\
     & case 2 & \Checkmark & \XSolidBrush  & 45.55 & 46.67 \\
     \rowcolor{pink}& default & \Checkmark & \Checkmark & \textbf{46.03} & \textbf{47.17} \\
    \bottomrule
    \end{tabular}%
  \label{tab:compress ablation}%
\end{table}%

\textbf{Embedding Size.}
As mentioned in Sec.~\ref{sec:preliminary}, the consumed bitrate of wavelength embeddings $\theta(\boldsymbol{e})$ and the decoder parameter $\theta(\mathcal{D})$ comprise the overall compressed bitrate. Given a certain rate constraint, it is necessary to make a trade-off between $\theta(\boldsymbol{e})$ and $\theta(\mathcal{D})$ for optimal compression efficiency. In Table~\ref{tab:embed size}, we evaluate the impact of embedding size on the reconstruction quality under a fixed total size of 0.5MB with 150 training epochs. We use size of $height \times width \times channel$ to denote a certain band embedding, thereby changing $\theta(\boldsymbol{e})$. It is suggested that the embedding size of $6 \times 3 \times 16$ with only 5\% bitrate consumption is the optimal choice.

\begin{table}[htbp]
% \footnotesize
  \centering
    \caption{Ablations on embedding size.}
    \begin{tabular}{ccc}
    \toprule
    Embedding Size & Embedding + Decoder & PSNR \\
    \midrule
    \midrule
    $12 \times 6 \times 16$  & 0.12 + 0.37 MB & 42.16 dB  \\
    $ \ \ 6 \times 3 \times 32$  & 0.06 + 0.43 MB & 42.58 dB \\
    \rowcolor{pink}\textbf{$ \ \ 6 \times 3 \times 16$}   & \textbf{0.03 + 0.47 MB} & \textbf{43.03 dB} \\
    $ 6 \times 3 \times 8$   & 0.02 + 0.48 MB & 42.82 dB \\
    $\ \ 4 \times 2 \times 16$  & 0.02 + 0.48 MB & 42.6 dB \\
    \bottomrule
    \end{tabular}%
  \label{tab:embed size}%
\end{table}%

\textbf{Content Angle Mapper.} Table~\ref{tab:loss ablation} presents a quantitative comparison under different loss functions. Introducing global supervision with CAM besides pixel-wise L1 loss yields improved HSI reconstruction quality. Furthermore, our proposed CAM demonstrates superior performance compared to the commonly used SSIM, indicating that CAM is more suitable for HSI reconstruction.

\begin{table}[htbp]
  \centering
  \caption{Ablation on the loss function.}
    \begin{tabular}{ccccc}
    \toprule
    \multicolumn{2}{c}{} & 0.5M  & 1M    & 2M \\
    \midrule
    \midrule
    \multicolumn{2}{c}{L1} & 43.78 & 44.66 & 45.97 \\
    \multicolumn{2}{c}{L1+SSIM} & 43.75 & 44.82 & 46.26 \\
    \rowcolor{pink}\multicolumn{2}{c}{L1+CAM} & \textbf{44.14} & \textbf{45.55} & \textbf{47.85} \\
    \bottomrule
    \end{tabular}%
  \label{tab:loss ablation}%
\end{table}%

\textbf{Classification.} 
Table~\ref{tab:classification ablation} illustrates the gradual transformation from using the original SF to our proposed method dedicated for compressed HSI. By adjusting the compressed reconstruction to adapt to classification task through ASW, and improving the generalization of classification model with augmented compressed HSI samples through ISI, we can achieve considerable accuracy improvement under high compression ratio.

\begin{table}[htbp]
% \footnotesize
  \centering
      \caption{\textbf{Ablation on proposed classification.}}
    \begin{tabular}{cccccc}
    \toprule
     & ASW   & ISI & \multicolumn{1}{c}{OA (\%)} & \multicolumn{1}{c}{AA (\%)} & \multicolumn{1}{c}{$\kappa$} \\
    \midrule
    \midrule
     case 1 (SF) & \XSolidBrush     & \XSolidBrush     & 79.15  & 84.27  & 0.7633  \\
     % \XSolidBrush     & \Checkmark     & 88.71  & 91.09  & 0.8706  \\
     case 2 & \Checkmark     & \XSolidBrush     & 84.75  & 89.88  & 0.8268  \\
     \rowcolor{pink!100}case 3 (Ours) & \Checkmark     & \Checkmark     & \textbf{87.03}  & \textbf{90.99} & \textbf{0.8519}  \\
    \bottomrule
    \end{tabular}%
  \label{tab:classification ablation}%
\end{table}%

\section{Conclusion}
\label{sec:Conclusion}
In this paper, we propose HINER, a novel neural representation for HSI. By explicitly embedding spectral wavelengths and introducing global CAM supervision, HINER effectively exploits correlation within and across spectral bands in HSI. Simultaneously, we thoroughly consider the degradation of downstream classification task induced by lossy compression. Through using {ASW} for classification-oriented reconstruction and {ISI} for data augmentation in training classification model, we effectively mitigate the task degradation. Experimental results demonstrate notable improvements in compression efficiency, model convergence, and classification accuracy compared to previous explorations. However, there is still plenty of room to improve our methodology. For instance, it falls short of the latest VVC in certain datasets, and still requires a few labels to supervise the classification model. These encourage us to pursue a more compact representation of HSIs and explore strategies involving soft label supervision in future work.

\begin{acks}
    This work was supported by Jiangsu Provincial Key Research and Development Program under Grant BE2022155. The authors would like to express their sincere gratitude to the Interdisciplinary Research Center for Future Intelligent Chips (Chip-X) and Yachen Foundation for their invaluable support.
\end{acks}

\balance

\bibliographystyle{ACM-Reference-Format}

% \bibliography{arxiv}

\clearpage
\appendix

\section{Classification on Compressed HSI}

\subsection{Adaptive Spectral Weighting (ASW)} \label{sec:saw}
\textbf{Architecture.}
{ASW} consists of two modules: WeightMLP $\mathcal{W}$ and ConvMLP $\mathcal{M}$, in which the output $\boldsymbol{I}_c \in \mathbb{R}^{N\times H\times W}$ maintains the original shape. Next, let's ignore the residual for simplicity, and this process can be abbreviated as:
\begin{align}
    \boldsymbol{I}_c = \mathcal{M}(\mathcal{W}(\hat{\boldsymbol{I}})).
    \label{eq:saw}
\end{align}

WeightMLP generates an n-dimensional vector $\boldsymbol{W} \in \mathbb{R}^{N\times 1}$ using a small MLP. The vector $\boldsymbol{W}$ is then utilized to weight the HSI spectral-wisely. The purpose of this step is to adaptively emphasize or de-emphasize certain spectral bands. Assuming $\hat{\boldsymbol{I}} \in\mathbb{R}^{N\times H \times W}$, the output $\boldsymbol{P}$ of WeightMLP can be written as:
\begin{align}
    \boldsymbol{P} = \hat{\boldsymbol{I}} \odot \boldsymbol{W} = 
        \begin{bmatrix} \hat{\boldsymbol{I}}_1 \\ \hat{\boldsymbol{I}}_2 \\ \dots  \\ \hat{\boldsymbol{I}}_n \end{bmatrix}
        \odot
        \begin{bmatrix} W_1 \\ W_2 \\ \dots  \\ W_n \end{bmatrix}
    = \begin{bmatrix} W_1\hat{\boldsymbol{I}}_1 \\ W_2\hat{\boldsymbol{I}}_2 \\ \dots  \\ W_n\hat{\boldsymbol{I}}_n \end{bmatrix}
    = \begin{bmatrix} \boldsymbol{P}_1 \\ \boldsymbol{P}_2 \\ \dots  \\ \boldsymbol{P}_n \end{bmatrix} \in \mathbb{R}^{N \times H \times W}.
    \label{eq:scalingnet}
\end{align}

Then $\boldsymbol{P}$ is passed to ConvMLP $\mathcal{M}$ comprising 1x1 conv to aggregate cross-spectral information. Let $\boldsymbol{A} \in \mathbb{R}^{N\times M \times 1 \times 1}$ and $\boldsymbol{B} \in \mathbb{R}^{M\times N \times 1 \times 1}$ represent the two convolution layers used in the $\mathcal{M}$, respectively:
\begin{align}
    \boldsymbol{A} &= \left[ \alpha_1, \alpha_2, \dots, \alpha_n \right]^T, \forall \alpha \in \mathbb{R}^{M \times 1}; \nonumber \\
    \boldsymbol{B} &= \left[ \beta_1, \beta_2, \dots, \beta_n \right], \forall \beta \in \mathbb{R}^{1 \times M}. 
    \label{eq:convmlp} 
\end{align}
Considering the $\lambda$-th band of output $\boldsymbol{I}_c$, it can be written as 
\begin{align}
    \boldsymbol{I}_c[\lambda,: ,:] = \alpha_1^T\beta_{\lambda}^T \boldsymbol{P}_1 + \alpha_2^T\beta_{\lambda}^T \boldsymbol{P}_2 + \dots
    \alpha_n^T\beta_{\lambda}^T \boldsymbol{P}_n
    \label{eq:convmlp out}
\end{align}
Combining Eq.~\ref{eq:saw}, it can be found that {ASW} first spectral-wisely re-weight the reconstructed HSI by multiplying learned vector $\boldsymbol{W}$, and then aggregate cross-spectral information.

\textbf{Optimization.} 
By employing {ASW}, the optimization of $\hat{\boldsymbol{I}}$ is converted into the optimization of network parameters. This conversion can be readily accomplished through gradient descent techniques. Then the input of classification network becomes the output of {ASW} $\boldsymbol{I}_c$, by which the the $\boldsymbol{u(\hat{\boldsymbol{I})}}=||\boldsymbol{I} - \hat{\boldsymbol{I}}||$ is translated to $\boldsymbol{u(\hat{\boldsymbol{I})}}=||\boldsymbol{I} - \boldsymbol{I_c}||$ to constrain classifier's input. We relax the constraint to prevent the necessity of introducing ground truth,
\begin{align}
    \boldsymbol{u}(\boldsymbol{I}) =
        ||\boldsymbol{I} - \boldsymbol{I_c}|| \approx
        ||\hat{\boldsymbol{I}} - \boldsymbol{I_c}|| 
        = ||\hat{\boldsymbol{I}} - \mathcal{SAW}(\hat{\boldsymbol{I}})||.
\end{align}

Given the condition $||\boldsymbol{I}-\hat{\boldsymbol{I}}|| < 10^{-3}$, this relaxation holds valid. Finally, our optimization objective can be expressed as the amalgamation of the classification loss $\mathcal{L}_C$ and the reconstruction loss $\mathcal{L}_R$, as described in Eq. (8) of the main paper:
\begin{align}
    \arg \min \mathcal{L}_C + \beta \cdot \mathcal{L}_R(\hat{\boldsymbol{I}}, \boldsymbol{I}_c)
\end{align}

\subsection{Implicit Spectral Interpolation (ISI)}
Data Augmentation has shown promise for training robust deep neural networks against unforeseen data bias or corruptions \cite{volpi2018generalizing, zhao2020maximum}. Intuitively, augmented samples encourage perturbing the underlying source distribution to enlarge predictive uncertainty of the current model, so that the generated perturbations can improve the model generalization during training. One intuitive manifestation of generalization is the flatness of the loss landscape. As described in the main paper, a flatter loss landscape, indicative of better generalization, exhibits relatively small loss changes under parameter perturbations, whereas a sharp loss landscape indicates otherwise.

We propose a simple yet effective strategy, Implicit Spectral Interpolation, to augment training samples, thereby facilitating improved performance on compressed HSI,
\begin{align}
    \mathcal{S} = \sum \mathcal{HINER} \left( \lambda + U(-\eta, \eta)\right),
\end{align}
where $U(-\eta, \eta)$ represents a uniform distribution that adds random variables to $\lambda$ to generate diverse reconstructed (perturbed) samples. In addition, we randomly disable and enable the spectral interpolation of the wavelengths in each forward pass, like \cite{wei2021qdrop}:
\begin{align}
    \eta = \left\{
    \begin{aligned}
        &0 \ \ \ \ \ \  with \ \  probability \ \ p \\
        &0.1 \ \ \  with \ \ probability \ \ 1-p
\end{aligned}
\right. .
\end{align}
Here we use $\eta=0.1$ and $p=0.5$.

\begin{figure*}[tbp]
\centering
\subfigure[Indian Pines]{
\label{fig:indian ori}
\includegraphics[scale=0.5]{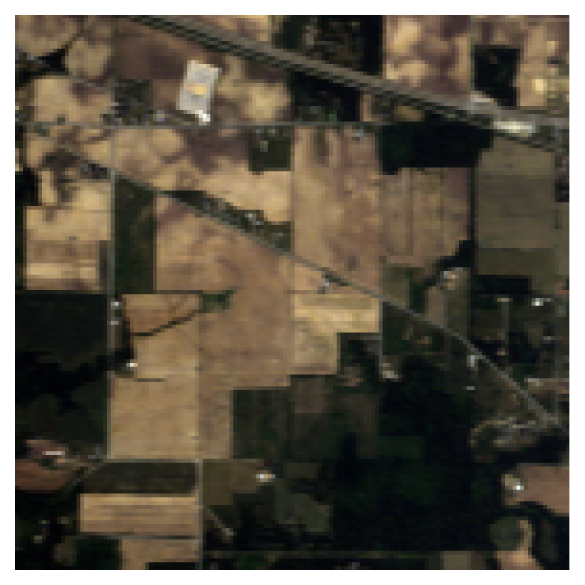}}
\subfigure[Pavia University]{
\label{fig:paviau ori}
\includegraphics[scale=0.5]{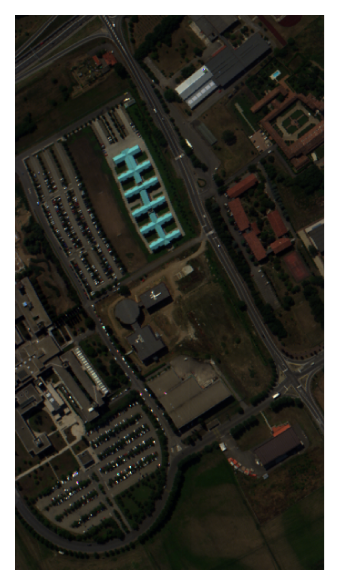}}
\subfigure[Pavia Centre]{
\label{fig:pavia ori}
\includegraphics[scale=0.5]{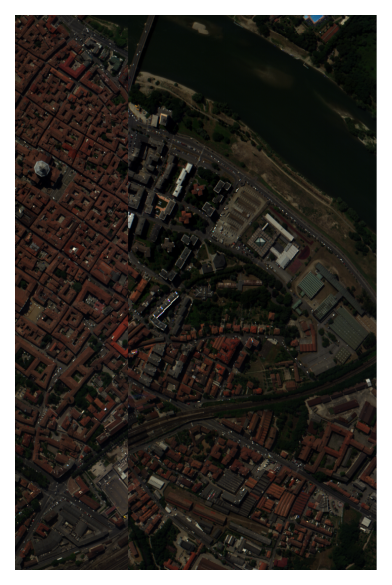}}
\subfigure[CHILD]{
\label{fig:child ori}
\includegraphics[scale=0.5]{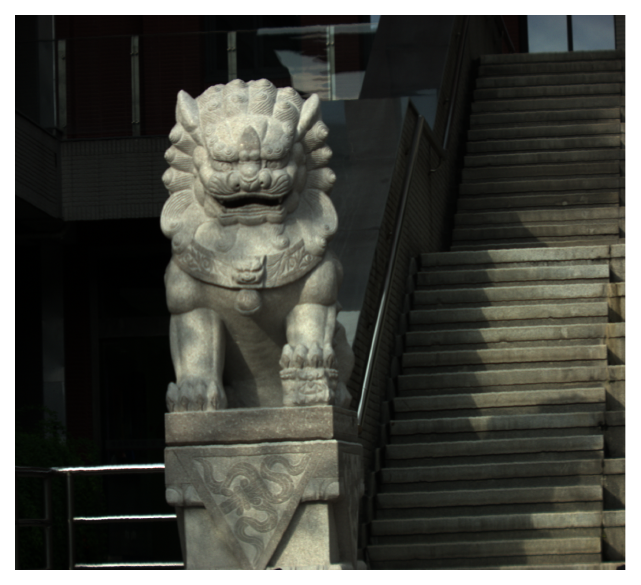}}
\caption{\textbf{Dataset visualization}}
\label{fig:ori}
\Description{}
\end{figure*}

\section{Experiments}

\subsection{Experimental Setup}

\subsubsection{Datasets}
\begin{itemize}
\item
\textbf{Indian Pines} is collected by the AVIRIS sensor over the Indian Pines Proving Ground in northwestern Indiana, used for compression and classification purposes. It consists a scene of 145$\times$145 pixels with 224 spectral bands spanning the wavelength range of 400-2500 nm. This scene is a subset of a larger scene. The Indian Pines scene predominantly consists of two-thirds agriculture and one-third forest or other perennial natural vegetation. Additionally, there are two major two-lane highways, a railroad line, and some low-density housing areas, along with other buildings and smaller roads. Sixteen classes are labeled (e.g., corn, grass, soybean, woods, and so on), with some classes being very rare (fewer than 100 samples for alfalfa or oats). After removing noisy bands, the number of bands is reduced to 200: [104-108], [150-163],220. Despite its limited size, this dataset serves as one of the main reference datasets in the community. A graphical representation of a sample from this dataset is presented in Fig. ~\ref{fig:indian ori}.

\item 
\textbf{Pavia University} is captured by the ROSIS sensor in Pavia, Northern Italy, with the purpose of compression and classification. The image dimensions are 610$\times$340 pixels, and it comprises 103 spectral bands. The image has been segmented into 9 distinct classes, including asphalt, meadows, gravel, trees, metal sheet, bare soil, bitumen, brick, and shadow.

\item 
{\textbf{Pavia Centre}} is a 1096$\times$715 pixels image, where the number of spectral bands is 102. The geometric resolution is 1.3 meters. Image differentiates 9 classes each, including water, trees, asphalt, self-blocking bricks, bitumen, tiles, shadows, meadows, and bare soil.

\item 
\textbf{CHILD}~\cite{huang2022high} comprises 141 HSI images captured by the PMVIS system, which measures 145 spectral samples ranging from 450 nm to 950 nm. The spatial resolution of each image is 960 $\times$ 1056 pixels. In this paper, we selected one HSI image from the dataset, named 20210803172558, for our experiment. Figure ~\ref{fig:child ori} shows its sample image.
\end{itemize}

Here Table \ref{tab:dataset} displays the training and testing datasets distribution in classification.

\begin{table}[htbp]
  \centering
  \caption{Land-cover classes of uesed Indian Pine and Pavia University datasets, with the standard training and testing distribution.}
    \begin{tabular}{ccccc}
    \toprule
          & classes & training & testing & spatial resolution \\
    \midrule
    \midrule
    Indian & 16    & 695 (3.3\%) & 9671  & 145x145 \\
    \midrule
    PaviaU & 9     & 3921 (1.9\%) & 40002 & 610x340 \\
    \bottomrule
    \end{tabular}%
  \label{tab:dataset}%
\end{table}%

\subsubsection{Evaluation Metrics}
\begin{itemize}
    \item {\textbf{PSNR}} (Peak Signal-to-Noise Ratio) quantifies the ratio of a signal to its noise, which calculates the ratio of the square of the maximum possible amplitude of the signal to the mean square error (MSE) in the signal. PSNR is employed as a measure of distortion in compression, where higher values correspond to better quality. The PSNR for an HSI with $N$ spectral bands can be formulated as:
    \begin{align}
        PSNR(I, \hat{I}) = \frac{1}{N} \sum_{i=1}^N 10\log_{10} \left( \frac{\max^2(I_i)}{MSE(I_i, \hat{I}_i)} \right)
        \label{eq:psnr}
    \end{align}
    
    % \item {\textbf{MS-SSIM}}: Indicator for measuring compression accuracy. Structural Similarity (SSIM) is a measure of the structural similarity of an image or video. It focuses more on image features perceived by the human eye, including brightness, contrast, and structure. SSIM evaluates quality by comparing the structural similarity between the original image and the processed image.SSIM takes values between 0 and 1, with 1 indicating the best similarity.
    
    \item {\textbf{bpppb}} (bits per pixel per band) is used to evaluate the consumption of compressed bitrate. For $I \in \mathbb{R}^{N\times H \times W}$, the bpppb is calculated as follows:
    \begin{align}
        bpppb = \frac{\theta(embeddings) \cdot b_e + \theta(decoder) \cdot b_d}{H \times W \times N}
    \end{align}
    where $\theta$ measures the parameters quantities and $b$ denotes the corresponding bit-width.

    \item {\textbf{CR}} (Compression Ratio) serves as a metric to quantify the compression effect, and it is defined as:
    \begin{align}
        CR = \frac{bpppb_{gt}}{bpppb_{compressed}}.
    \end{align}
    
    \item {\textbf{OA}} (Overall Accuracy) is employed to measure the overall classification accuracy. OA is calculated as the number of correctly categorized samples divided by the total sample size.
    \item {\textbf{AA}} (Average Accuracy) refers to the mean value of classification accuracy across all classes. It involves calculating the accuracy of each individual category and then averaging the accuracies of all categories.
    \item {{$\boldsymbol{\kappa}$}} (kappa coefficient) serves as a statistical measure of consistency between the classification maps and the ground truth. The $\kappa$ ranges from -1 to 1, where 1 signifies perfect consistency, 0 indicates stochastic consistency, and -1 implies complete inconsistency, where a higher $\kappa$ signifies better performance of the model.
\end{itemize}

\subsubsection{Implementation}
\begin{itemize}

\item \textbf{HINER.} 
In addition to the specifications outlined in the main paper, we employ a quantization bit-width of 8 bits for our experiments. Furthermore, for positional encoding, we set $b=1.25$ and $l=80$.

\item \textbf{FHNeRF and Rezasoltani.} 
FHNeRF and Rezasoltani are two state-of-the-art methods in the implicit neural representation of HSI, which take the original pixel coordinates as input and use $sine$ activations. Given that there are no publicly accessible source codes, we faithfully reproduce them. We use a 5-layer/15-layer perceptron and change the hidden dimension to build models of different sizes for FHNeRF~\cite{zhang2024compressing}/ Rezasoltani~\cite{rezasoltani2023hyperspectral_v0}, respectively. Both methods are trained for 15000 iterations with Adam optimizer~\cite{kingma2014adam} using a learning rate cosine descent strategy.

\item \textbf{JPEG2000.} 
For JPEG2000 compression, we utilize OpenJPEG to independently encode each spectral band. Initially, we transform the original HSI into individual raw files, with each file corresponding to a spectral band. Subsequently, we compress and decompress each raw file using OpenJPEG. After the decompression process, we convert the reconstructed raw files back into the MAT (matlab) format. This facilitates the comparison between the reconstructed data and the original data, enabling the computation of PSNR.

\item \textbf{VVC.} For VVC compression, we initially convert a MAT file into individual PNG files, with each PNG file corresponding to a spectral band. These PNG files are then merged into a YUV file, comprising a sequence of 'frames' at consecutive wavelengths. Subsequently, we perform compression using the VTM tool on the YUV file. However, due to VTM's lack of support for compressing 16-bit YUV files, we utilize 8-bit YUV files instead. After compression and subsequent decompression, we obtain the reconstructed YUV file. Next, we employ ffmpeg to convert the YUV file back into PNG files. The subsequent steps are akin to the JPEG2000 process, where we combine the individual PNG files back into a single MAT data format and compare the results with the original data to compute the PSNR.
\end{itemize}

\subsection{Encoding Complexity}
In Sec. 4.2 of the main paper, we have shown that {HINER} is faster than pixel-wise {FHNeRF} and {Rezasoltan} in encoding. Here, we further evaluate the image encoding speed compared to HNeRV, as shown in Table~\ref{tab:encoding time}. As observed, {HINER} achieves a higher speed compared to HNeRV, partly due to our encoder having fewer parameters. Additionally, after positional encoding, only a small input vector $\in \mathbb{R}^{1\times 160}$ needs to be processed by the MLP. This is smaller than the image matrix, e.g., $\in \mathbb{R}^{720 \times 360}$ in Pavia University, requiring multiple down-sampling operations with convolution. Consequently, our encoder has lower encoding complexity and better compression performance.
\begin{table}[htbp]
\caption{Encoding time comparison.}
  \centering
    \begin{tabular}{ccccc}
    \toprule
    \multirow{2}[2]{*}{Method} & \multirow{2}[2]{*}{Encoder Size} & \multicolumn{3}{c}{Model Size (MB)} \\
          &       & 0.2  & 0.5 & 1.5 \\
    \midrule
    \midrule
    \rowcolor{pink}\textbf{{HINER}}  & \textbf{0.12 MB}  & \textbf{480s}  & \textbf{500s}  & \textbf{790s} \\
    HNeRV & 0.22 MB  & 570s  & 620s  & 850s \\
    \bottomrule
    \end{tabular}%
  \label{tab:encoding time}%
\end{table}%

\begin{figure}[tb]
    \centering
    \subfigure[Training]{
    \includegraphics[scale=0.71]{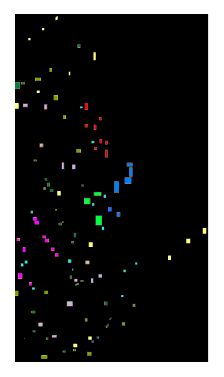}}
    \subfigure[Testing]{
    \includegraphics[scale=0.71]{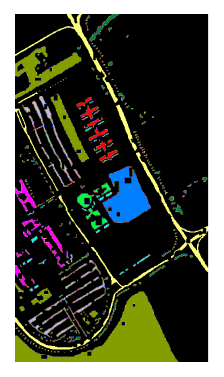}}
    \subfigure[Result]{
    \includegraphics[scale=0.71]{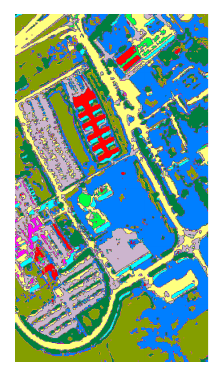}}
    \caption{Classification map obtained by our model on the Pavia University dataset}
    \Description{}
    \label{fig:class label}
\end{figure}

\subsection{Classification on Compressed HSI} 
Here, we present additional results regarding classification on compressed HSI samples. In Fig.\ref{fig:class label}, we visualize the spatial distribution of the training and testing sets, along with the classification map. Additionally, Table\ref{tab:indian class} and Table~\ref{tab:pavia class} exhibit quantitative performance at various compression ratios (CRs). For our method, all compressed HSIs with different CRs are evaluated using the same classification model trained at a CR of 28/109 for Indian Pine/ Pavia University datasets. In contrast, the SF method is re-trained for each CR to achieve the best performance. It is evident from the results that our method demonstrates superior performance in all cases, showcasing high robustness across various compression ratios.

\begin{table}[t]
% \footnotesize
  \centering
      \caption{Quantitative performance of the Indian Pines.}
    \begin{tabular}{lcccc}
    \toprule
    Method & \multicolumn{1}{c}{CR} & \multicolumn{1}{c}{OA (\%)} & \multicolumn{1}{c}{AA (\%)} & \multicolumn{1}{c}{$\kappa$} \\
    \midrule
    \midrule
    %2-D CNN & $\times1$ & 75.89  & 86.64  & 0.7281  \\
    %RNN  & $\times1$ & 70.66  & 76.37  & 0.6673  \\
    %miniGCN & $\times1$ & 75.11  & 78.03  & 0.7164  \\
    %ViT  & $\times1$ & 71.86  & 78.97  & 0.6804  \\
    % SF (pixel) & 16 & 78.55  & 84.68  & 0.7554  \\
    % FuNet-C & 79.89  & \textbf{89.35}  & 0.7716  \\
    \rowcolor{blue!20}SF & $\times1$ & 81.86  & 87.81 & 0.7919  \\
%    \textbf{Ours} & 16 & \textbf{88.34}  & \textbf{93.46} & \textbf{0.8671}  \\
    % \midrule
    % SF(pixel) $\star$ & 0.22 & 76.64 & 83.03 & 0.7361 \\
    % SF(patch) $\star$ & 0.22 & 80.61 & 85.9 & 0.7784 \\
    % \textbf{Ours} $\star$  & 0.22 & \textbf{85.63} & \textbf{89.95} & \textbf{0.8362} \\
    \midrule
    % SF (pixel) $\star$ & 0.58 & 74.15  & 81.49  & 0.7075  \\
    SF $^\clubsuit$ & $\times71$ & 80.61  & 85.90  & 0.7784  \\
    \rowcolor{pink}\textbf{Ours} $^\clubsuit$ & $\times71$ & \textbf{86.54}  & \textbf{91.17}  & \textbf{0.8465}  \\
    \midrule
    % SF (pixel) $\star$ & 0.58 & 74.15  & 81.49  & 0.7075  \\
    SF $^\clubsuit$ & $\times28$ & 79.15  & 84.27  & 0.7633  \\
    \rowcolor{pink}\textbf{Ours} $^\clubsuit$ & $\times28$ & \textbf{87.03}  & \textbf{90.99}  & \textbf{0.8519}  \\
    \midrule
    % SF (pixel) $\star$ & 0.58 & 74.15  & 81.49  & 0.7075  \\
    SF $^\clubsuit$ & $\times13$ & 79.23  & 86.73  & 0.7651  \\
    \rowcolor{pink}\textbf{Ours} $^\clubsuit$ & $\times13$ & \textbf{86.45}  & \textbf{90.94}  & \textbf{0.8457}  \\
    \bottomrule
    \end{tabular}%
  \label{tab:indian class}%
\end{table}%

\begin{table}[t]
% \footnotesize
  \centering
      \caption{\textbf{Quantitative performance of the Pavia University.}}
    \begin{tabular}{lcccc}
    \toprule
    Method & \multicolumn{1}{c}{CR} & \multicolumn{1}{c}{OA (\%)} & \multicolumn{1}{c}{AA (\%)} & \multicolumn{1}{c}{$\kappa$} \\
    \midrule
    \midrule
    %2-D CNN & $\times1$ & 86.05  & 88.99  & 0.8187  \\
    %RNN  & $\times1$ & 77.13  & 84.29  & 0.7101  \\
    %miniGCN  &$\times1$ & 79.79  & 85.07  & 0.7367  \\
    %ViT  & $\times1$ & 76.99  & 80.22  & 0.7010  \\
    % SF (pixel) & 16 & 87.94  & 87.47  & 0.8358  \\
    \rowcolor{blue!20}SF & $\times1$ & 91.07  & 90.20 & 0.8805  \\
    % \textbf{Ours} & 16 & \textbf{92.39}  & \textbf{91.36} & \textbf{0.8977}  \\
    % FuNet-C & \textbf{92.20}  & \textbf{89.66}  & \textbf{0.8951}  \\
    % \midrule
    % SF (pixel) $\star$  & 0.05 & 77.89  & 82.04  & 0.7164  \\
    % SF (patch) $\star$ & 0.05 & 85.78  & 88.35  & 0.8149  \\
    % \textbf{Ours} $\star$  & 0.05 & \textbf{91.63}  & \textbf{88.87}  & \textbf{0.8875}  \\
    \midrule
    % SF (pixel) $\star$  & 0.15 & 79.94  & 83.69  & 0.7431  \\
    SF$^\clubsuit$ & $\times109$ & 86.29  & 87.89  & 0.8203  \\
    \rowcolor{pink} \textbf{Ours}$^\clubsuit$  & $\times109$ & \textbf{ 88.93 }  & \textbf{ 88.96 }  & \textbf{ 0.8529 }  \\
    \midrule
    SF$^\clubsuit$  &$\times54$  & 86.75  & 88.77  & 0.8249  \\
    % SF (patch) $\star$ & 0.29 & 86.75  & 88.77  & 0.8249  \\
    \rowcolor{pink} \textbf{Ours}$^\clubsuit$  &$\times54$  & \textbf{88.55}  & \textbf{89.36}  & \textbf{0.8484}  \\
    \midrule
    SF$^\clubsuit$  &$\times35$  & 86.13  & 89.09  & 0.8189  \\
    % SF (patch) $\star$ & 0.29 & 86.75  & 88.77  & 0.8249  \\
    \rowcolor{pink} \textbf{Ours}$^\clubsuit$  &$\times35$  & \textbf{88.20}  & \textbf{89.27}  & \textbf{0.8438}  \\
    \bottomrule
    \end{tabular}%
  \label{tab:pavia class}%
\end{table}%

An interesting observation is that a lower compression ratio may not result in better accuracy. This phenomenon is consistent with previous works~\cite{wei2021effects, garcia2010impact, zabala2011effects} that for certain compression techniques, a higher CR may not significantly degrade the performance of pixel-based classification as the homogenization effect increases the similarity among pixels of the same area. In addition, land-cover type is also believed to be one of the factors as compression also has different effects on classification results of different land-cover types~\cite{zhai2008new}.

\subsection{Ablation Studies}
\subsubsection{\textbf{Positional Encoding}}
\begin{figure}[htbp]
    \centering
    \includegraphics[scale=0.55]{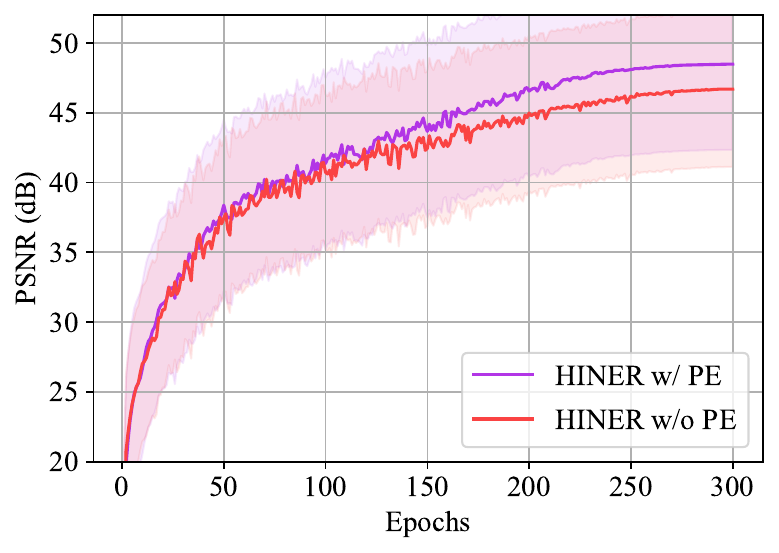}
    \caption{Regression curve of HINER w/ \& w/o PE.}
    \label{fig:pe}
    \Description{}
\end{figure}

As discussed in Section 3.2, MLPs are susceptible to the well-known spectral bias~\cite{xie2023diner, rahaman2019spectral}, wherein they tend to learn low frequency components of the signal. Thus, directly inputting the wavelength $\lambda$ into the encoder without positional encoding would lead to the network's incapacity to adequately capture high-frequency variation~\cite{mildenhall2021nerf, tancik2020fourier}. We illustrate this phenomenon with the regression curve shown in Figure~\ref{fig:pe}. Initially, during the earlier epochs, the performance of {HINER w/o PE} exhibits a similar regression performance with HINER w/ PE, indicating comparable capability in learning low-frequency components of the signal. However, as the epochs progress, the gap widens, highlighting the superior efficiency of positional encoding in capturing high-frequency information.

\subsubsection{\textbf{Reconstruction Loss}} 
Here, we present an ablation study concerning the $\gamma$ in Eq. (4) of the main paper:
\begin{align}
    \mathcal{L}_{R} = \underbrace{\sum_{n=1}^{N} || \hat{I}_n - I_n ||}_{L1 \ \ loss} 
    + {\color{red}\boldsymbol{\gamma}} \cdot
    \underbrace{\sum_{n=1}^{N} \frac{180}{\pi} \arccos \left( \frac{\Vec{\hat{I}}_n^T \cdot \Vec{I}_n}{\Vert \Vec{\hat{I}}_n^T \Vert_2 \Vert \Vec{I}_n \Vert_2}  \right)}_{CAM\ \ },
\end{align}
As depicted in Table~\ref{tab:cam_beta}, we set $\gamma=0.01$ in our experiments.

\begin{table}[htbp]
  \centering
  \caption{Ablations on coefficient $\gamma$ between L1 loss and CAM.}
    \begin{tabular}{cc}
    \toprule
    $\gamma$  & PSNR \\
    \midrule
    \midrule
    $0.005$ & 44.11 \\
    $0.001$ & 43.93 \\
    \rowcolor{pink}\textbf{$0.01$} & \textbf{44.14} \\
    $0.1$ & 43.87 \\
    \bottomrule
    \end{tabular}%
  \label{tab:cam_beta}%
\end{table}%

\subsubsection{\textbf{Adaptive Spectral Weighting}} We conduct thorough experiments from next two aspects.

\textbf{Optimization Objective.} 
As described in Sec. 3.3 of the main paper and Sec.~\ref{sec:saw} of the supplementary material, the optimization objective of ASW is formulated as:
\begin{align}
    \arg \min \mathcal{L}_C + \beta \cdot \mathcal{L}_R,
\end{align}
where $\mathcal{L}_R$ is introduced to constrain the input of the classifier (also the output of {ASW}) in the neighborhood of the ground truth. As illustrated in Table~\ref{tab:beta}, when $\mathcal{L}_R$ is omitted (i.e., $\beta=0$), there is a notable decrease in accuracy. This phenomenon also corroborates the validity of our theoretical analysis, i.e., \emph{{for downstream classification on compressed HSI, task accuracy is not only related to the classification loss but also to the reconstruction fidelity}}. Ultimately, we set $\beta=2.5$ to achieve a balance between these two losses.

\begin{table}[htbp]
  \centering
  \caption{Ablations on reconstruction loss.}
    \begin{tabular}{cccc}
    \toprule
    $\beta$  & OA(\%) & AA(\%) & $\kappa$ \\
    \midrule
    \midrule
    5     & 82.3  & 87.85 & 0.7979 \\
    \rowcolor{pink}2.5   & \textbf{87.03} & \textbf{90.99} & \textbf{0.8519} \\
    1.4   & 84.88 & 90.57 & 0.8282 \\
    0.5   & 83.88 & 88.6  & 0.8166 \\
    \rowcolor{blue!20}0     & 81.5  & 85.09 & 0.789 \\
    \bottomrule
    \end{tabular}%
  \label{tab:beta}%
\end{table}%

\textbf{Classification-Oriented Reconstruction.}
Additionally, we conduct ablation experiments to examine the effect of adding ASW before the classifier, as shown in Table~\ref{tab:saw psnr}. The inclusion of ASW results in PSNR decrease of the inputted reconstructed HSI of the classifier but an obvious improvement in classification accuracy. This suggests that {ASW} is able to adaptively weight HSI under the supervision of classification loss, thereby facilitating the translation of reconstruction from perceived visual quality to classification accuracy.

% Table generated by Excel2LaTeX from sheet 'Sheet4'
\begin{table}[htbp]
  \centering
  \caption{Ablations on ASW.}
    \begin{tabular}{lcccc}
    \toprule
          & \multicolumn{1}{l}{PSNR} & \multicolumn{1}{l}{OA (\%)} & \multicolumn{1}{l}{AA (\%)} & \multicolumn{1}{l}{$\kappa$} \\
    \midrule
    \midrule
    w/o ASW & \cellcolor{pink}\textbf{44.25} & \cellcolor{blue!20}79.15 & \cellcolor{blue!20}84.27 & \cellcolor{blue!20}0.7633 \\
    \midrule
    w/ ASW & \cellcolor{blue!20}34.71 & \cellcolor{pink}\textbf{84.06} & \cellcolor{pink}\textbf{88.24} & \cellcolor{pink}\textbf{0.8187} \\
    \bottomrule
    \end{tabular}%
  \label{tab:saw psnr}%
\end{table}%

\subsubsection{\textbf{Random uniform variables in ISI}}
In Sec. 3.3 of the main paper, we implement Implicit Spectral Interpolation (ISI) by introducing random variables on wavelengths:
\begin{align}
    \mathcal{S} = \sum \mathcal{HINER} \left( \lambda + U(-\eta, \eta)\right),
\end{align}
where $U(-\eta, \eta)$ represents a uniform distribution that adds random variables to $\lambda$. When trained with $\mathcal{S}$, the classification network exhibits improved generalization and reduced accuracy degradation on compressed HSI. It is crucial to note that ground truth HSI is not introduced during training for ISI. Table~\ref{tab:ISI} shows the impact of different $\eta$ (here the $\eta$ is not normalized). ISI proves to be a robust method across various $\eta$ values. For consistency, we set $\eta=0.1$ as the default setting.

\begin{table}[htbp]
  \centering
  \caption{Ablations on uniform perturbation $\eta$.}
    \begin{tabular}{cccc}
    \toprule
    \multicolumn{1}{c}{$\eta$} & \multicolumn{1}{c}{OA(\%)} & \multicolumn{1}{c}{AA(\%)} & \multicolumn{1}{c}{$\kappa$} \\
    \midrule
    \midrule
    0.05  & \cellcolor{orange!50}\textbf{87.13} & 90.06 & \cellcolor{blue!20}\textbf{0.8526} \\
    \rowcolor{pink}0.1   & 87.03 & 90.99 & 0.8519 \\
    0.15  & 85.68 & 90.9  & 0.8369 \\
    0.2   & 85.93 & 91.29 & 0.8399 \\
    0.4   & 86.88 & \cellcolor{yellow!80}\textbf{91.37} & 0.8502 \\
    \bottomrule
    \end{tabular}%
  \label{tab:ISI}%
\end{table}%

\end{document}